\documentclass[11pt]{article}
\usepackage[utf8]{inputenc}
\usepackage{setspace}
\usepackage[T1]{fontenc}
\usepackage{amsmath}
\usepackage{amsfonts}
\usepackage{amssymb}
\usepackage{graphicx}
\usepackage[a4paper,left=2cm,right=2cm,top=2cm,bottom=2cm]{geometry}
\usepackage{scrlayer-scrpage}
\usepackage{hyperref, newfloat, caption, subcaption}
\usepackage[english]{babel}
\usepackage[round]{natbib}
\usepackage{listings}
\usepackage{lineno}
\usepackage{xcolor}

\newtheorem{theorem}{Theorem}
\newtheorem{lemma}[theorem]{Lemma}
        
\hypersetup{
	colorlinks = true,
	urlcolor=blue,
	linkcolor=blue,
	citecolor=blue
}
\usepackage{natbib}
\usepackage{enumitem}

\newcommand{\ks}{\kappa_s }
\newcommand{\kw}{\kappa_w }

\newcommand{\vl}{c }

\newcommand{\por}{\boldsymbol{\theta} }
\newcommand{\porvec}{(\theta_1, \theta_2,... \theta_d) }
\newcommand{\porj}{\boldsymbol{\theta}^{(j)} }
\newcommand{\porjj}{\boldsymbol{\theta}^{(j-1)} }
\newcommand{\pormu}{\boldsymbol{\mu_{\theta}}}
\newcommand{\porsigma}{\boldsymbol{\Sigma_{\theta}}}
\newcommand{\porkappa}{C_{\theta}}
\newcommand{\pormuf}{\mu_{\theta}}

\newcommand{\data}{\boldsymbol{y} }
\newcommand{\datavec}{(y_1, y_2,...,y_T) }
\newcommand{\dataRV}{\boldsymbol{Y} }

\newcommand{\datanoise}{\boldsymbol{\varepsilon_{\mathcal{O}}}}

\newcommand{\lat}{\boldsymbol{X} }

\newcommand{\latvec}{(X_1,X_2,...,X_L) }
\newcommand{\latsmall}{\boldsymbol{x} }

\newcommand{\ppe}{\boldsymbol{\varepsilon_{\mathcal{P}}}}

\newcommand{\ppekappa}{C_{P}}
\newcommand{\ppesigma}{\boldsymbol{\Sigma_{P}}}

\newcommand{\perm}{\boldsymbol{\kappa}}

\newcommand{\muIS}{\boldsymbol{\mu_{IS}}}
\newcommand{\sigmaIS}{\boldsymbol{\Sigma_{IS}}}

\newcommand{\Zpor}{\boldsymbol{Z}}

\newcommand{\Zx}{\boldsymbol{Z_{P}}}
\newcommand{\Zppe}{\boldsymbol{Z_{P}}}

\let\oldequation\equation
\let\oldendequation\endequation
\renewenvironment{equation}
  {\linenomathNonumbers\oldequation}
  {\oldendequation\endlinenomath}
\let\oldalign\align
\let\oldendalign\endalign
\renewenvironment{align}
  {\linenomathNonumbers\oldalign}
  {\oldendalign\endlinenomath}
  
 \newcommand{\specialcell}[2][c]{
  \begin{tabular}
  [#1]{@{}c@{}}#2\end{tabular}}

\title{Lithological Tomography with the \\ Correlated Pseudo-Marginal Method}
\author{L. Friedli\footnote{Institute of Earth Sciences, University of Lausanne, Switzerland.}, N. Linde$^\ast$, D. Ginsbourger\footnote{Institute of Mathematical Statistics and Actuarial Science, University of Bern, Switzerland.}, A. Doucet\footnote{Department of Statistics, Oxford University, United Kingdom.}}

\begin{document}

\maketitle
\linespread{1.5}

Summary: We consider lithological tomography in which the posterior distribution of (hydro)geological parameters of interest is inferred from geophysical data by treating the intermediate geophysical properties as latent variables. In such a latent variable model, one needs to estimate the intractable likelihood of the (hydro)geological parameters given the geophysical data. The pseudo-marginal method is an adaptation of the Metropolis--Hastings algorithm in which an unbiased approximation of this likelihood is obtained by Monte Carlo averaging over samples from, in this setting, the noisy petrophysical relationship linking (hydro)geological and geophysical properties. To make the method practical in data-rich geophysical settings with low noise levels, we demonstrate that the Monte Carlo sampling must rely on importance sampling distributions that well approximate the posterior distribution of petrophysical scatter around the sampled (hydro)geological parameter field. To achieve a suitable acceptance rate, we rely both on (1) the correlated pseudo-marginal method, which correlates the samples used in the proposed and current states of the Markov chain, and (2) a model proposal scheme that preserves the prior distribution. As a synthetic test example, we infer porosity fields using crosshole ground-penetrating radar (GPR) first-arrival travel times. We use a (50~×~50)-dimensional pixel-based parameterization of the multi-Gaussian porosity field with known statistical parameters, resulting in a parameter space of high dimension. We demonstrate that the correlated pseudo-marginal method with our proposed importance sampling and prior-preserving proposal scheme outperforms current state-of-the-art methods in both linear and non-linear settings by greatly enhancing the posterior exploration. \\


\section{Introduction}
Geophysical investigations are rarely performed with the sole aim of inferring distributed subsurface models of geophysical properties. Rather, the underlying motivation is often to gain knowledge and constraints on other properties (e.g., permeability, clay fraction or mineral composition) and state variables (e.g., water saturation, salinity, temperature) of interest. Geophysical inverse theory has traditionally focused on assessing the resolution and uncertainty of inferred geophysical properties (e.g., \citeauthor{parker} \citeyear{parker}; \citeauthor{menke} \citeyear{menke}; \citeauthor{tarantola} \citeyear{tarantola}; \citeauthor{aster} \citeyear{aster}), while interpretation procedures in terms of properties or state variables of interest have received less attention. This is changing in hydrogeophysics (\citeauthor{binley} \citeyear{binley}), for instance, where it is now well-established that dedicated inversion approaches are needed when using geophysical data to gain knowledge about hydrogeological properties and state variables (e.g., \citeauthor{kowalski} \citeyear{kowalski}). For example, when inferring hydraulic conductivity by observing geophysical observables sensitive to water content or salinity during a tracer test experiment (\citeauthor{ldoetsch} \citeyear{ldoetsch}). However, these considerations have general validity and relevance for exploration and more fundamental geophysical studies. In a mantle context, for instance, one example concerns the inference of thermo-chemical constraints from seismological observations as reviewed by \citet{zunino}.\\

Multiple inversion frameworks have been proposed that combine hydrogeological and geophysical data in order to build predictive hydrogeological models (e.g., \citeauthor{ferre} \citeyear{ferre}; \citeauthor{ldoetsch} \citeyear{ldoetsch}). A critical aspect of such frameworks relates to how geophysical properties (sensed by geophysical data) are linked to hydrogeological target properties and variables of interest through petrophysical (rock physics) relationships. \citet{brunetti} distinguish between three sources of uncertainty related to petrophysical relationships: model uncertainty, parameter uncertainty and prediction uncertainty. While the first two refer to uncertainty in the choice of the appropriate petrophysical model and its parameter values, the latter is related to scatter and bias around the calibrated petrophysical model. In hydrogeophysical inversion studies targeting hydrogeological properties or state variables of interest, we note that the petrophysical relationship is often assumed to be perfect (deterministic) with known or unknown parameter values (e.g., \citeauthor{lochbuhler} \citeyear{lochbuhler}; \citeauthor{kowalski} \citeyear{kowalski}). However, ignoring petrophysical prediction uncertainty and its spatial correlation patterns results in bias, too narrow uncertainty bounds and overly variable hydrogeological parameter estimates (\citeauthor{brunetti} \citeyear{brunetti}). Unfortunately, analytical solutions to such inverse problems are available only when considering linear forward models and petrophysical relationships under the assumption of Gaussian distributions (\citeauthor{tarantola} \citeyear{tarantola}; \citeauthor{bosch2004} \citeyear{bosch2004}). Geophysical applications, however, often involve non-linear physics and non-linear petrophysical relationships (e.g., \citeauthor{mavko} \citeyear{mavko}).\\
	
Inversion approaches that account for petrophysical prediction uncertainty are often based on a two-step procedure: geophysical properties are first estimated using deterministic gradient-based inversions and then converted into parameters of interest using uncertain petrophysical relationships (e.g., \citeauthor{chen2001} \citeyear{chen2001}; \citeauthor{mukerji} \citeyear{mukerji}; \citeauthor{gonzalez} \citeyear{gonzalez}; \citeauthor{grana} \citeyear{grana}; \citeauthor{shahraeeni} \citeyear{shahraeeni}). The results of such a two-step approach can be misleading if neglecting the spatially-varying and typically much lower resolution of smoothness-constrained geophysical inversion models compared with the scale at which petrophysical relationships are developed (core or borehole logging scale) (\citeauthor{daylewis} \citeyear{daylewis}). Furthermore, with such an approach it is next to impossible to ensure that the geophysical inversion accounts for the prior constraints on the (hydro)geological target variable (\citeauthor{ferre} \citeyear{ferre}) and physical constraints such as conservation of mass, continuity and momentum. Moreover, for a deterministic inversion setting, \citet{bosch2004} showed that with a non-linear petrophysical relation, the two-step approach is an inherent approximation (\citeauthor{bosch2004} \citeyear{bosch2004}). \\
	
As an alternative to the two-step approach, coupled inversions directly target hydrogeological properties by inversion of geophysical data (e.g., \citeauthor{hinnell} \citeyear{hinnell}; \citeauthor{kowalski} \citeyear{kowalski}). They are often formulated within a Bayesian framework whereby one seeks to characterize the posterior probability density function (PDF) of hydrogeological parameters $\por$ given geophysical data~$\data$. Since it is often impossible to sample directly from the posterior PDF $p(\por | \data)$ of interest, Markov chain Monte Carlo (MCMC) methods, such as the Metropolis--Hastings method (MH; \citeauthor{hastings} \citeyear{hastings}; \citeauthor{metropolis} \citeyear{metropolis}), are used. Since the intermediate variable, the geophysical property $\lat$, connecting observations and target variables is unobservable (latent), one speaks of a latent variable model. In this study, we consider a setup where the latent geophysical property is given by $\lat = \mathcal{F}(\por) + \ppe$, with $\por \mapsto \mathcal{F}(\por)$ representing the deterministic component of a petrophysical relationship and $\ppe$ the petrophysical prediction error. Assuming an integrable and centered petrophysical prediction error $\ppe$, $\mathcal{F}(\por)$ stands for the expected value of the latent variable $\lat$. The geophysical data is given by $\dataRV = \mathcal{G}(\lat) + \datanoise$ with $\latsmall \mapsto \mathcal{G}(\latsmall)$ denoting the geophysical forward solver and $\datanoise$ describing the observational noise. \\

For a latent variable model as the one described above, the likelihood of observing the geophysical data given the proposed hydrogeological parameters, $p(\data | \por) = \int p(\data, \latsmall | \por) d \latsmall$, is often intractable. In the present context, this implies that the integral has an unknown or non-existing analytical form, which makes the direct implementation of the MH and related algorithms impossible. One way to circumvent this difficulty is to instead infer the joint posterior PDF~$(\por, \latsmall) \mapsto p(\por, \latsmall | \data)$ of the hydrogeological and geophysical parameters from which $p(\por|\data)$ is readily obtained by marginalization. Lithological tomography as introduced by \citet{bosch1999} pioneered such an approach to estimate the joint posterior by combining geophysical data, geological prior knowledge and uncertain petrophysical relationships. Within lithological tomography, pairs of the target and latent variables are proposed using marginal sampling of $\por$ and conditional sampling of $\lat$. Then, these pairs are accepted or rejected with $p(\data|\por, \latsmall)$, used in the acceptance ratio of the MH algorithm (where $p(\data|\por, \latsmall)=p(\data|\latsmall)$ is valid for our latent variable model). In \citet{bosch1999}, the conditional PDF $p(\latsmall|\por)$ to sample $\lat$ is given by a multivariate Gaussian distribution based on a suitable petrophysical relationship. In practice, this is achieved by adding brute force Monte Carlo realizations of the petrophysical prediction error $\ppe$ to the output of $\mathcal{F}(\por)$ at each iteration of the MCMC chain (i.e., \citeauthor{bosch2007} \citeyear{bosch2007}). \citet{linde2017} suggest that such an implementation is inefficient when considering large geophysical datasets with high signal-to-noise ratios and significant petrophysical uncertainty. The reason is that brute force Monte Carlo sampling of the petrophysical prediction error using $p(\latsmall|\por)$ induces high variability in the values taken by the likelihood function $p(\data|\por, \latsmall)$, even for the same $\por$, which could lead to prohibitively low acceptance rates even in the limiting case when the MCMC model proposal scale for $\por$ goes to zero. \\

\citet{brunetti} proposed an alternative approach to sample from the joint posterior PDF $p(\por, \latsmall | \data)$. In their method referred to herein as full inversion, the petrophysical prediction error $\ppe$ is parameterized and treated as the other unknowns within the MH algorithm. That is, the MH proposal mechanism draws new realizations of both the target variable $\por$ and the petrophysical prediction error $\ppe$, which combined also lead to a realization of the latent variable $\lat$ used to calculate the likelihood function $p(\data|\por, \latsmall)$. \citet{brunetti} presented a convincing performance of the full inversion approach with clear improvements in efficiency compared with the original formulation of lithological tomography by \citet{bosch1999}. Nonetheless, the full inversion method suffers from high dimensionality, and the strong (posterior) correlation between $\ppe$ and $\por$ makes standard MCMC inversions inefficient (e.g., \citeauthor{CPM} \citeyear{CPM}). \\	
	
In this study, we evaluate an inversion method targeting directly the marginal posterior $p(\por|\data)$ by approximating the intractable likelihood $p(\data | \por) = \int p(\data | \por, \latsmall) p(\latsmall|\por) d \latsmall$. In the pseudo-marginal (PM) method introduced by \citet{beaumont} and studied by \citet{PM}, the true likelihood is replaced with a non-negative unbiased estimator resulting in a MH algorithm sampling the same target distribution as when using the true likelihood. In their work, \citet{beaumont} and \citet{PM} use an unbiased likelihood estimator based on Monte Carlo averaging over samples of the latent variable. In our setting with the latent variable $\lat = \mathcal{F}(\por) + \ppe$, we note that the original lithological tomography approach of \citet{bosch1999} is closely related to the pseudo-marginal method. In the original lithological tomography method targeting the joint posterior PDF $p(\por, \latsmall | \data)$, the MCMC chains store the conditional draws of the latent variables together with the target variables, and the target posterior PDF $p(\por |\data)$ is obtained by marginalization. The PM method applied with one draw of the latent variable leads to equivalent results in terms of the marginal posterior PDF. In the PM method, the draws of the latent variable are not stored but only used to estimate the likelihood $p(\data | \por)$. Using only one sample of the latent variable in the PM method typically leads to impractically-low acceptance rates due to the high variability of the ratio of log-likelihood estimators. To achieve an efficient algorithm, the standard deviation of the log-likelihood estimator needs to be around 1.2-1.5 (\citeauthor{PMBiomet} \citeyear{PMBiomet}), which is ensured by increasing the number of samples and applying importance sampling. schemes. In the context of state-space models, the number of Monte Carlo samples used in the likelihood estimator needs to increase linearly with the number of observations, which becomes impractical in data-rich applications (\citeauthor{CPM} \citeyear{CPM}). To obtain low-variance log-likelihood ratio approximations with a smaller number of Monte Carlo samples, \citet{CPM} introduced the correlated pseudo-marginal (CPM) method by which the draws of latent variables used in the denominator and numerator in the likelihood ratio are correlated. Both the PM and CPM methods are general in that they allow for non-linear and non-Gaussian assumptions, but their implementation and applicability in data-rich high-dimensional geophysical settings remain untested.  \\

Inferring hundreds or thousands of parameters with a MH algorithm is challenging as the number of iterations needed for convergence grows with the number of target parameters (e.g., \citeauthor{robert2018} \citeyear{robert2018}). To ensure adequate performance in such settings, it is crucial to equip the algorithm with a well-working proposal scheme. In the context of Gaussian random fields with high dimension, \citet{cotter2013} demonstrated that standard random walk MCMC algorithms leads to strong dependence on the discretization of the target field and highly inefficient algorithms. Their proposed solution lies in preserving the prior PDF within the proposal scheme such that the acceptance probability of model proposals only depends on the likelihood ratio. This type of proposal schemes was explored in geophysics by \citet{mosegaard}, in what is often referred to as the extended Metropolis algorithm. In a high-dimensional target space, the extended Metropolis approach still needs an efficient model proposal scheme (\citeauthor{ruggeri} \citeyear{ruggeri}). Following \citet{brunetti}, we use the adaptive multi-chain algorithm DREAM$_{(ZS)}$ (DiffeRential Evolution Adaptive Metropolis using an archive of past states) by \citet{dream_zs}, which is widely used in various geophysical inversion studies (e.g., \citeauthor{d1} \citeyear{d1}; \citeauthor{d3} \citeyear{d3}; \citeauthor{d2} \citeyear{d2}). We adapt herein the DREAM$_{(ZS)}$'s formulation in order to accommodate prior-preserving model proposals. 	\\

As an exemplary problem, we consider inference of high-dimensional multi-Gaussian porosity fields using crosshole ground-penetrating radar (GPR) first-arrival travel times. We consider both a linear straight-ray solver, to enable comparisons with analytical solutions, and a more physically-based non-linear eikonal solver. We compare the results obtained by our prior-sampling-based proposal and importance-sampling-based implementation of the (correlated) pseudo-marginal method with standard model proposals and without importance sampling. Furthermore, we compare against the original lithological tomography formulation, full inversion and MCMC inversions that simply ignore the presence of petrophysical prediction uncertainty. With these examples, we will demonstrate that our implementation of the CPM method is outperforming the other inversion methods by greatly enhancing the posterior exploration. \\

This paper is structured as follows. Section \ref{2} introduces the methodology by discussing Bayesian inference in the context of high-dimensional settings, presenting the inversion approaches considered and the tools employed for performance assessment. Section \ref{3} presents the two test examples with linear and non-linear physics. The results and wider implications are discussed in Section \ref{4}, followed by conclusions in Section \ref{5}.  

\section{Methodology}
\label{2} 

The methodology section starts by introducing the considered latent variable model (Section \ref{lvm}), followed by general considerations concerning Bayesian inference and MCMC in high-dimensional settings (Section \ref{bi}). The correlated pseudo-marginal method and our IS procedure are introduced in Section \ref{CPM} and baseline methods used for comparative purposes are presented in Section \ref{baseline}. Finally, Section \ref{assessment} presents the performance assessment metrics used to evaluate the results.  

\subsection{Latent variable model}
\label{lvm}
We consider a latent variable model where the unobservable variable $\lat=\latvec$ is related to the $d$ target parameters~$\por=\porvec$ and the $T$ measurements $\data=\datavec$.  We write 
\begin{equation}
	\dataRV = \mathcal{G}(\lat) + \datanoise  = \mathcal{G}(\mathcal{F}(\por) + \ppe) + \datanoise,
\end{equation}
for $\mathcal{G}: \mathbb{R}^L \rightarrow \mathbb{R}^T$ and $\mathcal{F}: \mathbb{R}^d \rightarrow \mathbb{R}^L$ with errors $\datanoise$ and $\ppe$. In our setting, $\boldsymbol{x} \mapsto \mathcal{G}(\boldsymbol{x})$ describes the physical forward solver with $\datanoise$ denoting the observational noise and $\por \mapsto \mathcal{F}(\por)$ represents the petrophysical relationship with $\ppe$ denoting the petrophysical prediction error (PPE). We assume both errors to be Gaussian such that the distribution of $\lat | \por$ can be represented with the PDF $p( \latsmall | \por)= \varphi_{L}(\latsmall; \mathcal{F}(\por), \boldsymbol{\Sigma_P})$ and the one of $\dataRV | \por, \lat$ with the PDF $p(\data | \por, \latsmall )=\varphi_{T}(\data; \mathcal{G}(\latsmall), \boldsymbol{\Sigma_{\dataRV}})$, with the notation $\varphi_{M}(\cdot; \boldsymbol{\mu}, \boldsymbol{\Sigma})$ denoting the PDF of a $M$-variate Normal distribution with mean $\boldsymbol{\mu}$ and covariance matrix $\boldsymbol{\Sigma}$.

\subsection{Bayesian Inference with Markov Chain Monte Carlo}
\label{bi}
	
In Bayes' theorem, the posterior probability density function~(PDF) $p(\por | \data )$ of the model parameters $\por$ given the measurements $\data$ is specified by
\begin{align}
	p(\por | \data ) = \frac{p(\por) p(\data | \por) }{p(\data)},
\end{align}
with the prior PDF~$p(\por)$ of the model parameters, the likelihood function~$p(\data | \por)$ and the evidence~$p(\data)$. Generally, there is no analytical form of the posterior PDF. If the posterior PDF can be evaluated pointwise up to a normalizing constant, MCMC methods can be used to generate posterior samples. The basic idea of MCMC algorithms is to construct a Markov chain with the posterior PDF of interest as its stationary distribution (see e.g., \citeauthor{MCstat} \citeyear{MCstat}). MCMC algorithms iteratively propose new values for the states of the Markov chain that are accepted or rejected with a prescribed probability. One foundational MCMC algorithm is Metropolis--Hastings (MH; \citeauthor{metropolis} \citeyear{metropolis}; \citeauthor{hastings} \citeyear{hastings}). It proceeds as follows at iteration $j$: First, using the model proposal density $q(\cdot| \porjj)$, a new set of states $\porj$ is proposed. Then, the acceptance probability, 
\begin{align}
		\label{MHr}
		\alpha_{MH} \left( \porjj, \porj \right) &= \min \biggl\{ 1, \frac{q(\porjj| \porj)p(\porj|\data)}{q(\porj|  \porjj)p(\porjj|\data)} \biggr\} 
		= \min \biggl\{ 1, \frac{q(\porjj| \porj)p(\porj)p(\data | \porj)}{q(\porj| \porjj)p(\porjj)p(\data|\porjj)} \biggr\},
\end{align}
is calculated and the proposed $\porj$ is accepted (if $\alpha_{MH}(\porjj, \porj) \geq V$) or rejected \linebreak (if $\alpha_{MH}(\porjj, \porj)~<~V$) on the basis of a draw of a uniformly distributed random variable $V \sim Unif([0,1])$. If the proposed $\porj$ is rejected, the old state of the chain is kept and $\porj = \porjj$. \\

Within the MH algorithm, we need to evaluate the likelihood function $\por \mapsto~p(\data | \por)$ in order to compute the acceptance probability. In our latent variable model (see Section \ref{lvm}), the likelihood is given by,
\begin{equation}
		\label{LHint}
		p(\data | \por) = \int p(\data | \por, \latsmall) p(\latsmall| \por) d \latsmall,
\end{equation}	
and the integral has generally no analytical form. In Sections \ref{CPM}, \ref{LithTom} and \ref{FI}, we present three methods to circumvent the difficulties of an intractable likelihood function. 

\subsubsection{Model parameterization and proposal scheme}	
\label{par}

We consider test examples targeting a Gaussian random field $GRF(\pormuf(\cdot), \porkappa(\cdot, \cdot))$ with known mean $\pormuf(\cdot)$ and covariance function $\porkappa(\cdot, \cdot)$. We parameterize the target field $\por$ using a regular 2D grid of size \textit{D} $\times$ \textit{D} (such that $d=D^2$ for the notation introduced in Section \ref{lvm}) with positions $\mathcal{B}~=~\{b_1,b_2,...,b_{D^2}\}$: 
\begin{equation}
		\por \sim \mathcal{N}_{D^2}(\pormu, \porsigma), \text{ with } \pormu = (\pormuf(g_i))_{1 \leq i \leq D^2} \text{ and } \porsigma = (\porkappa(g_i, g_j))_{1 \leq i,j \leq D^2},
\end{equation}
with $\mathcal{N}_{D^2}(\boldsymbol{\mu}, \boldsymbol{\Sigma})$ denoting the $D^2$-variate normal distribution with mean $\boldsymbol{\mu}$ and covariance matrix~$\boldsymbol{\Sigma}$. We use a high-dimensional pixel-based parameterization of the target field, $\por~=~\pormu~+~\porsigma^{1/2}~\Zpor$, where $\Zpor$ is a $D^2$-dimensional random vector consisting of $i.i.d.$ standard-normal distributed variables. To infer the target field, we need to estimate the $\Zpor$-variables. Similar to \citet{ruggeri}, we do not apply any further dimensionality reduction of the parameter space beyond the discretization (in contrast with, for instance, \citeauthor{brunetti} (\citeyear{brunetti}) who used the dimensionality reduction approach of \citeauthor{Generate_GRF} (\citeyear{Generate_GRF})). This is done to avoid  distorted posterior PDF estimates that may arise in response to a reduction of the parameter space. Furthermore, we seek to evaluate performance in a challenging high-dimensional setting with thousands of unknowns. \\
	
When inferring model parameters with the MH algorithm, it is crucial to choose the model proposal scale well. If the model proposal steps are too large, the acceptance rate is low and the Markov chain needs many iterations until convergence. If the step-width is too small, the exploration of the parameter space is very slow and the Markov chain will similarly need many iterations until convergence (see Section \ref{assessment} for the assessment of convergence). To deal with this challenge of tuning the proposal scale of each model parameter, we use the adaptive multi-chain algorithm DREAM$_{(ZS)}$ (DiffeRential Evolution Adaptive Metropolis using an archive of past states) by \citet{dream_zs} for which details can be found in Appendix \ref{appendix_dream}. \\
	
MCMC algorithms generally suffer from the curse of dimensionality as the number of iterations needed for convergence increases with the number of target parameters (e.g., \citeauthor{robert2018} \citeyear{robert2018}). In the context of Gaussian random fields, \citet{cotter2013} show that MCMC methods based on standard random walk proposals lead to strong dependencies on the discretization of the target field and to inefficient algorithms when employed in high dimensions. For a given proposal scale, refining the grid representing the random field leads to a decreasing acceptance rate with zero as the limiting value for an infinite number of unknowns. To make MCMC algorithms robust to discretization and maintain a reasonable stepsize when inferring thousands of unknowns, they propose model proposal schemes such as the pCN (preconditioned Crank-Nicholson) that preserve the prior PDF. For a target variable $\boldsymbol{Z}$ with a Standard-Normal prior, the proposal of a standard random walk method is given by $\boldsymbol{Z}^{(j)} = \boldsymbol{Z}^{(j-1)} + \gamma \zeta$, with $\gamma$ being the step size and $\zeta \sim \mathcal{N}(0,1)$, respectively. Instead, the pCN proposal scheme uses $\boldsymbol{Z}^{(j)} = \sqrt{1-\gamma^2} \boldsymbol{Z}^{(j-1)} + \gamma \zeta$, ensuring that $\boldsymbol{Z}^{(j)}$ remains standard-normally distributed. \citet{cotter2013} show that proposal schemes preserving the prior PDF lead to (1) algorithms that mix more rapidly and (2) the convergence being insensitive to the discretization of the target field. We note that the idea of defining a model proposal scheme preserving the prior distribution was proposed more than 25 years ago in geophysics by \citet{mosegaard}. This approach is often referred to as the extended Metropolis algorithm and has mainly been explored in the context of inversion with complex geostatistical prior models (a detailed description of the method can be found in \citeauthor{hansen2012} (\citeyear{hansen2012})). Defining a proposal density $q(\cdot| \porjj)$ such that the MCMC algorithm samples the prior PDF in the absence of data implies that $\frac{q(\porjj| \porj)}{q(\porj| \porjj)} = \frac{p(\porjj)}{p(\porj)}$ holds true, with the implication that the MH acceptance-ratio of Equation~(\ref{MHr}) is reduced to the likelihood ratio, 
	\begin{align}
		\label{MHr2}
		\alpha_{MH} \left( \porjj, \porj \right) = \min \biggl\{ 1, \frac{p \big( \data | \porj \big) }{p \big( \data|\porjj \big) } \biggr\}.
	\end{align}
The extended Metropolis approach still needs an efficient model proposal scheme (\citeauthor{ruggeri} \citeyear{ruggeri}), which is why we use DREAM$_{(ZS)}$ in this work. In the case of a Gaussian-distributed prior, the standard DREAM$_{(ZS)}$ proposal scheme does not generate samples that preserve the prior distribution. In order to adapt extended Metropolis to DREAM$_{(ZS)}$, we rely on a transformation of the variables to the Uniform space (details in Appendix \ref{appendix_dream}). This transformation makes it possible to create a proposal mechanism which unites (1) the efficiency of the DREAM$_{(ZS)}$ proposals with (2) the robustness of the prior-preserving proposals. In what follows, our proposal scheme using the uniform transform will be referred to as prior-sampling DREAM$_{(ZS)}$ proposals, while the the standard proposal scheme of DREAM$_{(ZS)}$ will be referred to as standard DREAM$_{(ZS)}$ proposals. We stress that both prior-sampling DREAM$_{(ZS)}$ and standard DREAM$_{(ZS)}$ target the same posterior PDF, but the former is expected to be more efficient.

\pagebreak
	\subsection{(Correlated) pseudo-marginal method}	
	\label{CPM}

	\subsubsection{Pseudo-marginal method}	
	\citet{beaumont} shows that a MH algorithm using a non-negative unbiased estimator of the likelihood samples the same target distribution as when using the true likelihood. He exploits this property by estimating the likelihood in Equation (\ref{LHint}) on the basis of Monte Carlo averaging over samples of the latent variable~$\lat$. \citet{PM} adopt this approach in their pseudo-marginal (PM) method and provide a theoretical analysis of the scheme. When one brute force Monte Carlo sample of the latent variable is drawn in each MCMC iteration without importance sampling (c.f., the original lithological tomography by \citeauthor{bosch1999} (\citeyear{bosch1999}); see Section \ref{LithTom}), the algorithm is likely to suffer from a low acceptance rate due to the high variability of the log-likelihood estimator. This is due to the fact that a likelihood estimator given by $p(\data | \por, \boldsymbol{X})$ takes very different values depending on the draw of the latent variable $\lat$, even for the same~$\por$. This occurs as the scatter ($\ppe$) has a strong effect on the data response, and hence, the likelihood. To improve the efficiency, \citet{beaumont} and \citet{PM} use many samples drawn by importance sampling (IS; e.g. \citeauthor{IS} \citeyear{IS}). Consequently, they propose the following unbiased estimator of the likelihood $p(\data|\por)$,
\begin{align}
	\label{pmsum1}
	\hat{p}_N(\data| \por) = \frac{1}{N} \sum\limits_{n=1}^{N} w(\data | \por, \boldsymbol{X}_n), \quad \text{with} \quad w(\data | \por, \boldsymbol{X}_n) = 
	\frac{ p(\data | \por, \boldsymbol{X}_n) p(\boldsymbol{X}_n| \por) }{ m(\boldsymbol{X}_n | \por) },
\end{align}
where $\boldsymbol{X}_n \overset{i.i.d}{\sim} m(\cdot | \por)$ for $n=1,2,...,N$ with $m(\cdot | \por)$ being the importance density function. More details about the importance sampling procedure will follow in Section \ref{S.IS}. 

\subsubsection{Correlated pseudo-marginal method}
\label{sec_cpm}
For the PM method to be efficient, the number of samples $N$ used in the likelihood estimator~(Eq. (\ref{pmsum1})) should be selected such that the variance of the log-likelihood ratio estimator is low enough (\citeauthor{PMBiomet} \citeyear{PMBiomet}). If it is too high, the algorithm will suffer from an impractically low acceptance rate. In the state-space model context, this implies that $N$ needs to scale linearly with $T$ leading to a computational cost of order $T^2$ at every MCMC iteration, which can be prohibitively expensive for large $T$ (\citeauthor{CPM} \citeyear{CPM}). To reduce the computational cost, \citet{CPM} introduced the correlated pseudo-marginal (CPM) method by which the draws of latent variables used in the denominator and numerator of the likelihood ratio estimators are correlated. The underlying idea is that the variance of a ratio of estimators is lower if they are positively correlated (\citeauthor{koop} \citeyear{koop}). Assuming that the latent variable $\lat$ is standard-normal distributed, the CPM method proposes (in iteration $j$) a realization of the $n$-th latent variable draw by means of pre-conditioned Crank-Nicholson proposals,
	\begin{equation}
		\label{corr}
	\boldsymbol{X}_n^{(j)} = \rho \boldsymbol{X}_n^{(j-1)} + \sqrt{1-\rho^2} \boldsymbol{\epsilon}, \text{ with } \rho \in (0,1) \text{ and } \boldsymbol{\epsilon} = (\epsilon_1, \epsilon_2,...,\epsilon_L ), \epsilon_i \overset{i.i.d.}{\sim} \mathcal{N}(0,1).
	\end{equation}
The assumption that the latent variable has a standard-normal distribution hardly limits the general applicability of the CPM method, since there exist transformations from numerous distributions that will allow proposals to act on Gaussian distributions (e.g. \citeauthor{chen2018} \citeyear{chen2018}; Section \ref{S.IS}). We stress that if the proposed $\porj$ with $\boldsymbol{X}_n^{(j)}$ is rejected by the CPM algorithm, we keep $\boldsymbol{X}_n^{(j)}=\boldsymbol{X}_n^{(j-1)}$ as for $\porj = \porjj$. \\
	
Compared to standard MCMC algorithms, the CPM method requires two additional parameters: the latent variable sample size $N$ and the correlation parameter~$\rho$. To achieve optimal performance, the parameters should be chosen such that the variance of the log-likelihood ratio estimator for a fixed target variable $\por$,
	\begin{equation}
	\label{loglhesti}
	 R=\log \left(\hat{p}_N^{(j)}(\data|\por) \right)  - \log \left( \hat{p}_N^{(j-1)}(\data|\por) \right),
	\end{equation}
takes values between 1.0 and 2.0 in regions with high probability mass (\citeauthor{CPM} \citeyear{CPM}). Here, $\hat{p}_N^{(j)}(\data|\por)$ and $\hat{p}_N^{(j-1)}(\data|\por)$ refer to the likelihood estimators (Eq. (\ref{pmsum1})) obtained with the accepted latent variable of iteration $j-1$ and the proposed (and not necessarily accepted) latent variable of iteration $j$, that is, the likelihood estimators used in the acceptance ratio of the MH algorithm. In order to choose the parameter values, we first fix the number of samples $N$ at a value that is smaller than the number of available parallel processors. Then, we evaluate different $\rho$ and estimate corresponding values of $Var(R)$ for a fixed $\por$ in a region with high posterior probability mass (e.g., chosen based on initial MCMC runs).

\subsubsection{Importance sampling procedure}
\label{S.IS}
For high-dimensional problems with large data sets exhibiting high signal-to-noise ratios, it is necessary to use importance sampling when drawing samples of latent variables to be used within the likelihood-estimator (Eq. (\ref{pmsum1})). This is a consequence of the integrand $p(\data | \por, \latsmall)$ in Equation~(\ref{LHint}) having a peak in a region of $\lat$ having small probability under $p(\latsmall| \por)$. Importance sampling proceeds by sampling from a so-called importance distribution given by the PDF $\latsmall \mapsto m(\latsmall | \por)$ that preferentially generates samples with high $p(\data | \por, \latsmall)p(\latsmall|\por)$. Furthermore, the support of the importance distribution must include all values $\latsmall$, for which $p(\data | \por, \latsmall)p(\latsmall| \por)>0$ (\citeauthor{IS} \citeyear{IS}). It holds,
\begin{equation}
	\int p(\data | \por, \latsmall) p(\latsmall| \por) d \latsmall = \int \frac{p(\data | \por, \latsmall) p(\latsmall| \por)}{m(\latsmall|\por)} m(\latsmall|\por) d \latsmall,
\end{equation}
leading to the unbiased importance sampling estimate of the likelihood given in Equation (\ref{pmsum1}). To ensure minimal variance of the estimator, we seek $\latsmall \mapsto m(\latsmall | \por)$ to be nearly proportional to $\latsmall~\mapsto~p(\data | \por, \latsmall)p(\latsmall| \por)$ as recalled in \citet{IS} referring to the results of \citet{kahn}. Since $p(\latsmall|\por, \data)~\propto~p(\data | \por, \latsmall)p(\latsmall| \por)$, it is sensible to base the importance density on $\latsmall \mapsto p(\latsmall|\por, \data)$.\\

Within a latent variable model with a non-linear physical forward solver (Section \ref{lvm}), we can not derive the exact expression for $p(\latsmall|\por, \data)$. Here, we derive local approximations of this posterior by relying on linearization. To do so, we use a linearization of the map $\boldsymbol{x} \mapsto \mathcal{G}(\boldsymbol{x})$ around $\latsmall_{lin} = \mathcal{F}(\por_{lin}) + \ppe_{lin}$ based on a first-order expansion, 
\begin{align}
\label{foe}
    \mathcal{G}( \latsmall) = \mathcal{G}( \latsmall_{lin} + \latsmall  - \latsmall_{lin}) \approx \mathcal{G}( \latsmall_{lin}) +  \boldsymbol{J}_{\latsmall_{lin}} (\latsmall  - \latsmall_{lin}),
\end{align} 
with $\boldsymbol{J}_{\latsmall_{lin}}$ being the Jacobian matrix of the forward solver corresponding to $\latsmall_{lin}$. Ideally, $\latsmall_{lin}$ should be given by a realization of the latent variable similar to the one the algorithm is currently exploring. By approximating $p(\data | \por, \latsmall)$ with $\widetilde{p}(\data | \por, \latsmall) = \varphi_{T}(\data; \mathcal{G}( \latsmall_{lin}) +  \boldsymbol{J}_{\latsmall_{lin}} (\latsmall  - \latsmall_{lin}), \boldsymbol{\Sigma_{\dataRV}})$ and, applying $p(\latsmall | \por) = \varphi_{L}(\latsmall; \mathcal{F}( \por), \boldsymbol{\Sigma_{P}})$ and the relationships between marginal and conditional Gaussians out of \citet{bishop} given in Appendix \ref{appendix2}, we get, 
\begin{align}
	\label{IS11}
	&\widetilde{p}(\latsmall | \por, \data) = \varphi_{L}(\latsmall; \muIS,\sigmaIS), \text{ with}  \\
	&	\muIS = \sigmaIS \left( \boldsymbol{J}_{\latsmall_{lin}}^T \boldsymbol{\Sigma_{\dataRV}}^{-1} \left(  
	\data - ( \mathcal{G}\left( \latsmall_{lin} \right) - \boldsymbol{J}_{\latsmall_{lin}} \latsmall_{lin} ) \right) 
	+ \ppesigma ^{-1} \mathcal{F}(\por) \right), \nonumber \\
	&\sigmaIS = (\ppesigma ^{-1} + \boldsymbol{J}_{\latsmall_{lin}}^T \boldsymbol{\Sigma_{\dataRV}}^{-1} \boldsymbol{J}_{\latsmall_{lin}})^{-1}, \nonumber
\end{align}
for an approximation of $p(\latsmall | \por, \data)$. To incorporate importance sampling within the CPM method, we need to correlate the draws of latent variables. To achieve this, we rely on the fact that a realization of the latent variable $\lat$ can be generated with $\muIS + \sigmaIS^{1/2} \Zx$, where $\Zx$ is standard Gaussian distributed in $\mathbb{R}^L$. Using this representation, we can correlate the (standard-normal distributed) $\Zx$-variables using Equation (\ref{corr}). 

\subsection{Baseline inversion methods}
\label{baseline}

We present now the inversion approaches used for comparison with the CPM method. These include a method ignoring the petrophysical prediction errors and two approaches (original formulation of lithological tomography without importance sampling and full inversion) accounting for the PPEs by inferring the joint posterior PDF~$p(\por, \latsmall | \data)$ of the target and latent variables. An overview of all inversion methods (including CPM) is given in Table \ref{tab:methods}. 
\begin{table}
\caption{\label{tab: table-name} Overview of the inversion methods applied on the latent variable model introduced in Section~\ref{lvm}; a box around a letter indicates that this parameter is saved as a target variable of the MH algorithm. For the proposal scheme we use both standard and prior-sampling DREAM$_{(ZS)}$ proposals for all methods.}
\begin{center}
\begin{small}
\setlength{\tabcolsep}{0pt}
\def\arraystretch{1.5}
\begin{tabular}{c c c c}
	 \specialcell{\textbf{Method}} &  \specialcell{\textbf{Proposal} \\ \textbf{ scheme}} & \specialcell{\textbf{Latent variable(s)}} & \specialcell{\textbf{Likelihood $\hat{p}(\data|\por)$}} \\
	\hline \hline
	\specialcell{\textbf{No PPE}: \\ Ignore PPE} &  \specialcell{$\boxed{\porj}$} & \specialcell{$\lat^{(j)} = \mathcal{F}(\porj)$} &
	\specialcell{$\varphi_T(\data; \mathcal{G}(\lat^{(j)}), \mathbf{\Sigma}_{\dataRV})$} \\
	\hline
	\specialcell{\textbf{Full inversion}: \\ Infer PPE} &  \specialcell{$\boxed{\porj,  \ppe^{(j)}}$} &
	\specialcell{$\lat^{(j)} = \mathcal{F}(\porj)+\ppe^{(j)}$} & \specialcell{
	$\varphi_T(\data; \mathcal{G}(\lat^{(j)}), \mathbf{\Sigma}_{\dataRV})$} \\
	\hline
	\specialcell{\textbf{LithTom}: \\ Infer PPE} &  \specialcell{$\boxed{\porj}$} & \specialcell{$\boxed{\lat^{(j)}} \sim \varphi_{L}(\cdot; \mathcal{F}(\porj), \mathbf{\Sigma}_P)$}& \specialcell{
	$\varphi_T(\data; \mathcal{G}(\lat^{(j)}), \mathbf{\Sigma}_{\dataRV})$}\\
	\hline
	\specialcell{\textbf{LithTom IS}: \\ Infer PPE} &  \specialcell{$\boxed{\porj}$} & \specialcell{$\boxed{\lat^{(j)}} \sim \varphi_{L}(\cdot; \muIS, \sigmaIS)$}& \specialcell{$\frac{ \varphi_T(\data; \mathcal{G}(\lat^{(j)}), \mathbf{\Sigma}_{\dataRV}) \varphi_{L}(\lat^{(j)}; \mathcal{F}(\porj), \mathbf{\Sigma}_P) }{ \varphi_{L}(\lat^{(j)}; \muIS, \sigmaIS) }$}\\
	\hline
	\specialcell{\textbf{(C)PM no IS}: \\ Sample out PPE} &  \specialcell{$\boxed{\porj}$}  &
	\specialcell{ $\lat^{(j)} = (\lat^{(j)}_1,...,\lat^{(j)}_N)$ \\
	$\lat^{(j)}_n \overset{i.i.d}{\sim} \varphi_{L}(\cdot; \mathcal{F}(\porj), \mathbf{\Sigma}_P)$ \\
	\textbf{CPM}: Correlation $\lat^{(j-1)}_n$} &
    \specialcell{$\frac{1}{N} \sum\limits_{n=1}^{N} \varphi_T(\data; \mathcal{G}(\lat^{(j)}_n), \mathbf{\Sigma}_{\dataRV})$} \\
	\hline
    \specialcell{\textbf{(C)PM IS}: \\ Sample out PPE} &  \specialcell{$\boxed{\porj}$} &
	\specialcell{ $\lat^{(j)} = (\lat^{(j)}_1,...,\lat^{(j)}_N)$ \\
	$\lat^{(j)}_n \overset{i.i.d}{\sim} \varphi_{L}(\cdot; \muIS, \sigmaIS)$ \\
	\textbf{CPM}: Correlation $\lat^{(j-1)}_n$}&
    \specialcell{$\frac{1}{N} \sum\limits_{n=1}^{N}\frac{ \varphi_T(\data; \mathcal{G}(\lat^{(j)}_n), \mathbf{\Sigma}_{\dataRV}) \varphi_{L}(\lat^{(j)}_n; \mathcal{F}(\porj), \mathbf{\Sigma}_P) }{ \varphi_{L}(\lat^{(j)}_n; \muIS, \sigmaIS) }$  } 
\end{tabular}
\end{small}
\end{center}
\label{tab:methods}
\end{table}

\subsubsection{Ignore petrophysical prediction errors}

This inversion method (no PPE) ignores the presence of petrophysical prediction errors in the MH algorithm. For the latent variable model introduced in Section \ref{lvm}, this results in an approximation of the likelihood function with the Gaussian PDF $\hat{p}(\data|\por) = \varphi_T(\data; \mathcal{G}(\mathcal{F}(\por)), \mathbf{\Sigma}_{\dataRV})$, where the forward response $\mathcal{G}(\mathcal{F}(\por))$ is simulated without accounting for PPEs. The method is included in the comparison as it is commonly used in practice as discussed by \citet{brunetti}.

\subsubsection{Lithological Tomography}
\label{LithTom}

One way to consider PPEs while circumventing the difficulty of an intractable likelihood function is to infer the joint posterior PDF~$(\por, \latsmall) \mapsto p(\por, \latsmall | \data)$ of the hydrogeological and geophysical parameters. Lithological tomography (\citeauthor{bosch1999} \citeyear{bosch1999}) pursues this strategy and uses a factorization of the joint posterior PDF as $p(\por, \latsmall | \data) \propto p(\por) p(\latsmall|\por) p(\data|\por, \latsmall)$, where $p(\data|\por, \latsmall) = p(\data|\latsmall)$ is valid for our setting. To sample from this posterior PDF, \citet{bosch1999} proceeds as follows: First, realizations from the joint prior of $\por$ and $\lat$ are created by marginal sampling of $\por$ and conditional sampling of $\lat$. Then, the pairs of model proposals are accepted or rejected with $p(\data|\latsmall)$, used in the acceptance ratio of the MH algorithm. In practice, this means that brute force Monte Carlo realizations (no importance sampling) of the petrophysical prediction error $\ppe$ are added to the output of the petrophysical relationship $\mathcal{F}(\por)$. For our latent variable model, this results in an approximation of the likelihood function with $\hat{p}(\data|\por) = \varphi_T(\data; \mathcal{G}(\latsmall), \mathbf{\Sigma}_{\dataRV})$, where the latent variable $\lat = \mathcal{F}(\por) + \ppe$ is obtained with a draw of $\ppe$ from the multivariate Gaussian with PDF $\varphi_{L}(\cdot; 0, \mathbf{\Sigma}_P)$.

\subsubsection{Full Inversion}	
\label{FI}
The full inversion approach infers the joint posterior PDF by treating the latent variables analogously to the other unknowns. In the context of our latent variable model~(Section~\ref{lvm}), this means that in iteration $j$ of the MH, not only a new $\porj$ but also a new $\ppe^{(j)}$ is proposed by the algorithm's proposal scheme. Then the likelihood function $p(\data|\por, \latsmall) = \varphi_T(\data; \mathcal{G}(\latsmall), \mathbf{\Sigma}_{\dataRV})$ is calculated using $\lat^{(j)} = \mathcal{F}(\porj)+\ppe^{(j)}$. \citet{brunetti} applied full inversion to infer porosity fields by inversion of crosshole GPR first-arrival travel times, that is, to a setting similar to ours. For the parametrization of the porosity field of interest, they used a spectral representation combined with the dimensionality reduction approach of \citet{Generate_GRF}.  \citet{brunetti} achieved convincing results and improvements compared to standard lithological tomography without importance sampling (Section \ref{LithTom}). Nevertheless, full inversion is expected to suffer from high dimensionality and strong correlation among the latent and target variables as the two sets of variables are treated as being independent within the proposal scheme (e.g., \citeauthor{CPM} \citeyear{CPM}). 
	
\subsection{Performance assessment}
\label{assessment}
To assess the performance of the different inversion approaches, we primarily focus on the exploration of the posterior PDF. The reason for this will become clear in the results section (Section \ref{3}). \\

To declare convergence, we use the $\hat{R}$-statistic of \citet{rstat} that compares the within-chain variance with the between-chain variance for the second half of the MCMC chains. The general convention is that convergence is declared once this statistic is smaller or equal to 1.2 for all model parameters. Since we deal with a high-dimensional parameter space with thousands of unknowns, we relax this condition slightly and declare convergence if 99 $\%$ of the parameters satisfy this criterion. When an algorithm is considered convergent, we compare the resulting posterior samples with those of the other approaches. \\

For the test case with linear physics in Section \ref{linear}, we compare the results with the analytical solution of the posterior PDF $p(\por|\data)$. For these comparisons, we use histograms and the Kullback--Leibler divergence (KL~-~divergence; \citeauthor{KLdiv} \citeyear{KLdiv}). The KL~-~divergence between two PDFs $p_1(\cdot)$ and $p_2(\cdot)$ is defined as,
\begin{equation}
	KL(p_1 || p_2) = \int p_1(x) \log \left( \frac{p_1(x)}{p_2(x)}\right) dx.
\end{equation}
To obtain the PDF of the estimated posterior, we can use the MCMC samples to either (1) make a kernel density estimate or to (2) estimate the mean and variance for a Gaussian approximation (\citeauthor{krueger} \citeyear{krueger}). Here we use the second option since the posterior is Gaussian. If the PDFs $p_1(\cdot)$ and $p_2(\cdot)$ are Gaussians with $p_1=\mathcal{N}(\mu_1,\sigma_1^2)$ and  $p_2=\mathcal{N}(\mu_2,\sigma_2^2)$, the expression of the KL-divergence reduces to,
\begin{equation}
	\label{KLG}
	KL(p_1 || p_2) = \log \left( \frac{\sigma_2}{\sigma_1}\right) +  \frac{\sigma_1^2 + (\mu_1-\mu_2)^2}{2 \sigma_2^2} - \frac{1}{2}.
\end{equation}
A KL-divergence of zero indicates that the two PDFs are equal and it increases as the distributions diverge from each other. \\

For the test example with non-linear physics in Section \ref{nonlinear}, there is no analytical solution to compare with. Hence, we compare the estimated posterior distribution with a single value (the known true porosity at each pixel). We achieve this by applying so-called scoring rules (\citeauthor{gneiting} \citeyear{gneiting}) assessing the accuracy of a predictive PDF $\por \mapsto \hat{p}(\por)$ with respect to a true value $\por$. Scoring rules are functions that assign a numerical score for each prediction-observation pair $(\hat{p},\por)$, with a smaller score indicating a better prediction. They assess both the statistical consistency between predictions and observations (calibration) and the sharpness of the prediction. We use the logarithmic score (logS; \citeauthor{good} \citeyear{good}) defined by $\text{logS}(\hat{p},\por) = -\log \hat{p}(\por)$ that is related to the Kullback--Leibler divergence (\citeauthor{gneiting} \citeyear{gneiting}). As for the linear case, we use the MCMC samples to obtain a Gaussian approximation of the estimated posterior PDF. The logarithmic score favours predictive PDFs under which the true value has high probability. We supplement this measure with two simpler ones: the number of pixels in which the true porosity value was in the range of the posterior samples and the standard deviation of the estimated posterior PDF. \\

We also consider the acceptance rates (AR) and the integrated autocorrelation time (IACT). We aim for an acceptance rate of 15$\%$~-~30$\%$ as proposed by \citet{DREAM}. The IACT of the chain $\{\por^{(j)}; j=1,2,...\}$ is defined as $1 + 2 \sum\limits_{l=0}^{\infty} Corr(\por^{(1)}, \por^{(1+l)})$. In practice, the estimated autocorrelation for large values of $l$ is noisy such that we need to truncate the sum. Following \citet{gelman}, we truncate the sum when two successive autocorrelation estimates are negative. We renounce from discussing the CPU time as it depends strongly on the chosen forward model and discretization as well as on other parameters pertaining to the computing equipment.

\section{Results}
\label{3}
	
We consider the problem of inferring the porosity distribution using crosshole GPR first-arrival travel times. We first address a test case with linear physics (straight-rays) to allow for comparison with analytical solutions and then one with non-linear physics (eikonal solver) to address a more challenging and physically-based setup. Our examples are synthetic and the water-saturated porosity field is described by a multi-Gaussian random field.

\subsection{Data and inversion setting}

\subsubsection{Synthetic data generation}
\label{data}
\begin{figure}
		\centering
		\begin{subfigure}{.325\linewidth}
			\centering
			\includegraphics[height=4.3cm]{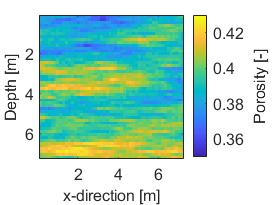}
			\caption{}
			\label{fig:setup1}
		\end{subfigure}
			\begin{subfigure}{.325\textwidth}
		\centering
		\includegraphics[height=4.3cm]{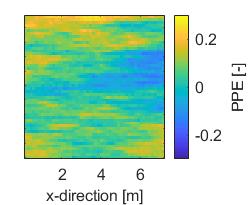}
		\caption{}
		\label{fig:setup2}
	\end{subfigure}
		\begin{subfigure}{.325\textwidth}
	\centering
	\includegraphics[height=4.3cm]{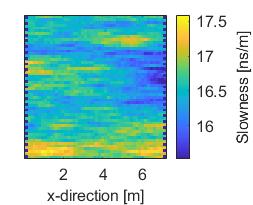}
	\caption{}
	\label{fig:setup3}
\end{subfigure}
\begin{subfigure}{.48\textwidth}
	\centering
	\includegraphics[width=1\linewidth]{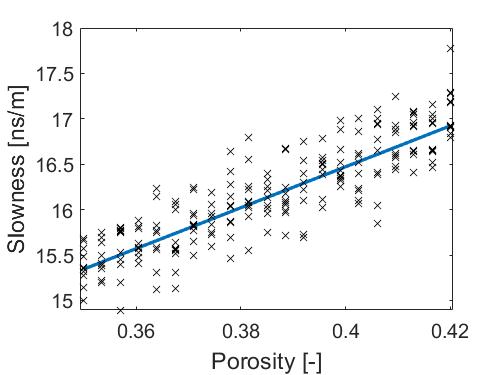}
	\caption{}
	\label{fig:setup7}
\end{subfigure}
\begin{subfigure}{.48\textwidth}
	\centering
	\includegraphics[width=1\linewidth]{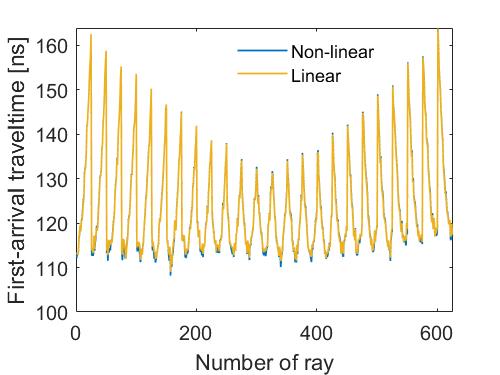}
	\caption{}
	\label{fig:setup6}
\end{subfigure}
		\caption{(a) Porosity field $ \por$, (b) PPE field $\ppe$, (c) slowness field $\latsmall$ with transmitter-receiver layout, (d) dependency of slowness on porosity obtained without (line) and with (scatter) PPE and (e) noise-contaminated first-arrival travel times $\data$ for the linear and the non-linear forward solver corresponding to the true synthetic model.}
		\label{fig:setup}
\end{figure}

Our considered subsurface domain is 7.2 m $\times$ 7.2 m and we use 25 equidistant GPR transmitters located on the left side and 25 receivers on the right side of the model domain, resulting in 625~first-arrival travel times. The transmitter-receiver layout is depicted in Figure~\ref{fig:setup3}. As introduced in Section~\ref{par}, we assume the porosity field to be a Gaussian random field $GRF(\pormuf(\cdot), \porkappa(\cdot, \cdot))$. We use $\pormuf(\cdot)=0.39$ and an exponential covariance function $\porkappa(\cdot, \cdot)$. For the latter, we use a sill of $2e^{-4}$ and geometric anisotropy where the main, horizontal direction has an integral scale of $4.5$ m and the integral scale ratio between the horizontal and vertical direction is $0.13$. We use a (50 $\times$ 50)-dimensional pixel-based parameterization of the porosity field; the true synthetically generated field is shown in Figure \ref{fig:setup1}. Note that porosity is a positive quantity bounded between zero and one while a Gaussian prior distribution has a full support. The Gaussian prior is used here to ensure an analytical solution in the linear physics case. Given the presented mean and the sill, it is extremely unlikely that a porosity value outside the physical boundaries is generated. In other settings, one could use a transform of the porosity (e.g., as in \citeauthor{bosch2004} \citeyear{bosch2004}) or choose a bounded distribution. \\
		
To predict the dielectric constant $\perm$, we use the complex refractive index model (CRIM; \citeauthor{crim} \citeyear{crim}),
\begin{equation}
		\sqrt{\perm} = \sqrt{\ks} + (\sqrt{\kw}-\sqrt{\ks}) \por,
		\end{equation}
where $\kw$ and $\ks$ are the dielectric constants of water [81] and mineral grains [5], respectively.  The resulting slowness field (which in our case is the latent variable $\lat$) depicted in Figure \ref{fig:setup3} is given by,
		\begin{equation}
		\label{velo}
		\latsmall = \sqrt{\vl^{-2}\perm} + \ppe = \frac{1}{c} \big( \sqrt{\ks} + (\sqrt{\kw}-\sqrt{\ks}) \por \big) + \ppe,
		\end{equation}
where $\vl$ is the speed of light in vacuum [0.3 m/ns]. This specifies the petrophysical relationship to be linear with $\por \mapsto \mathcal{F}(\por) =  \frac{1}{c} \left( \sqrt{\ks} + (\sqrt{\kw}-\sqrt{\ks}) \por \right)$. We add a petrophysical prediction error~(PPE) $\ppe$ that is a realization of a centred GRF over a regular 2D grid of size 50 $\times$ 50. We are assuming that the PPE field (depicted in Figure \ref{fig:setup2}) has an exponential covariance function $\ppekappa (\cdot, \cdot)$ with a sill of $2.1e^{-2}$ and the same correlation structure as the porosity field. The dependency of the slowness on the value of the porosity and the PPE is indicated in Figure \ref{fig:setup7}. Finally, the resulting 625 GPR first~-~arrival travel times are calculated with (i) a linear (straight-ray) forward solver referred to as $\mathcal{G}_{s}$ and (ii) a non-linear (eikonal) forward solver referred to as $\mathcal{G}_{e}$ (the \textit{time2D} solver of \citeauthor{time2D} (\citeyear{time2D})), such that,
		\begin{equation}
		\data = \mathcal{G}(\latsmall) + \datanoise,
		\end{equation}
with $i.i.d.$ centered normal observational noise $ \datanoise$ with standard deviation of 1~ns. The two sets of traveltimes are depicted in Figure \ref{fig:setup6}.
		
\subsubsection{Inversion settings and prior assumptions}
All considered inversion methods (Sections \ref{CPM} and \ref{baseline}) are implemented with prior-sampling and standard DREAM$_{(ZS)}$ proposals using the same parameter settings of the DREAM$_{(ZS)}$ algorithm with four MCMC chains running in parallel. For the prior on porosity, we use the Gaussian PDF $p(\por) = \varphi_{2500}(\por; \pormu, \porsigma)$ assuming the mean $\pormu$ and covariance structure $\porsigma$ to be known (the same values as for the data generation). Using a pixel-based parameterization of the field, we infer the 2500-dimensional vector $\Zpor$ defining the porosity by $\por = \pormu + \porsigma^{1/2} \Zpor$, with $\Zpor$ having a multivariate standard-normal prior PDF. The full inversion has to estimate another 2500~$\Zppe$-variables for the PPE field leading to a total of 5000 inferred parameters. For the PPE $\ppe$ we also use a Gaussian prior PDF $p(\ppe) = \varphi_{2500}(\por; 0, \ppesigma)$ with known covariance structure $\ppesigma$, leading to a Gaussian prior PDF for the slowness field (for fixed porosity) given by $p(\latsmall | \por) = \varphi_{2500}(\latsmall; \mathcal{F}(\por), \ppesigma)$. For the likelihood function, we assume that the 625-dimensional vector describing the observational noise $\datanoise$ has a Gaussian distribution with zero mean and diagonal covariance matrix $\boldsymbol{\Sigma_{\dataRV}}$; the standard deviation is assumed to be 1 ns as in the data generation process. 

\subsection{Linear physics}
\label{linear}
	To enable comparisons of the inferred posterior PDFs with the analytical solution for $p(\por | \data)$, we first consider the case of linear physics. Then,
		\begin{equation}
		\data = \mathcal{G}_{s}(\latsmall) + \datanoise = \boldsymbol{J_{s}} \latsmall + \datanoise,
		\end{equation}
	with $\boldsymbol{J_{s}}$ being the Jacobian (i.e., forward operator) of the linear forward solver. The analytical posterior PDF can be derived as detailed in Appendix \ref{appendix2}. Figure \ref{fig:truepost1} shows the posterior mean and Figures \ref{fig:truepost2} - \ref{fig:truepost4} depict three draws from the posterior distribution. \\
	
	\begin{figure}
			\centering
			\begin{subfigure}{.23\textwidth}
				\centering
			\includegraphics[height=4cm]{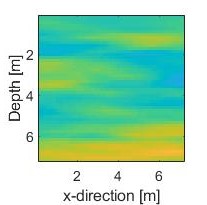}
				\caption{}
				\label{fig:truepost1}
			\end{subfigure}
			\begin{subfigure}{.21\textwidth}
				\centering
			\includegraphics[height=4cm]{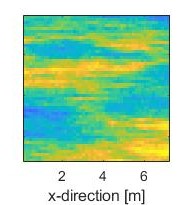}
				\caption{}
				\label{fig:truepost2}
			\end{subfigure}
			\begin{subfigure}{.21\textwidth}
				\centering
			\includegraphics[height=4cm]{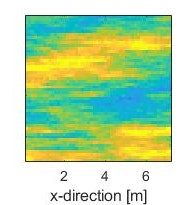}
				\caption{}
				\label{fig:truepost3}
			\end{subfigure}
			\begin{subfigure}{.29\textwidth}
				\centering
         	\includegraphics[height=4cm]{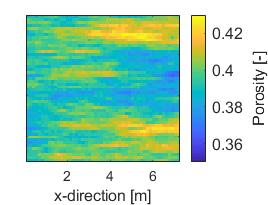}
				\caption{}
				\label{fig:truepost4}
			\end{subfigure}
			\caption{(a) Analytical posterior mean of $p(\por|\data)$ for the linear test example and (b) - (d) three realizations of the analytical posterior distribution.}
			\label{fig:truepost}
	\end{figure}

When employing the PM and CPM method in this setting of large datasets with low noise, it is crucial to use a well-chosen importance sampling for the latent variable. As introduced in Section \ref{S.IS}, it is sensible to use $\latsmall \mapsto p(\latsmall|\por, \data)$ as a basis for the importance density. As long as we are in the linear Gaussian case, we can derive the analytical expression for this posterior (Appendix \ref{appendix2}), resulting in a zero-variance importance sampling density (\citeauthor{IS} \citeyear{IS}). Since it then does not make sense to use multiple importance density samples (the importance weights are constant), we combine in this linear case importance sampling with PM using \textit{N}=1 (original lithological tomography algorithm enhanced with importance sampling that we will hereafter refer to as LithTom IS). We note that using the exact formula for the importance sampling corresponds to having access to the exact likelihood $p(\por|\data)$. The use of larger \textit{N} is considered in Section \ref{nonlinear} for the case of non-linear physics. This linear setting for which analytical solutions are available serves mainly (1) to demonstrate the necessity of a well-working importance sampling distribution, (2) to investigate the exploration capabilities of MCMC-based inversion approaches that estimate the intractable likelihood using Monte Carlo samples (lithological tomography, PM and CPM methods) and (3) to compare the performances of the prior-sampling and standard DREAM$_{(ZS)}$ proposal mechanisms. \\

\begin{figure}[h]	
	\begin{flushleft}
		\begin{subfigure}{.225\textwidth}
			\centering
		\includegraphics[height=4cm]{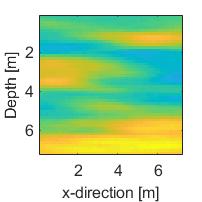}
			\caption{}
			\label{fig:res_lin1}
		\end{subfigure}
		\begin{subfigure}{.21\textwidth}
			\centering
		\includegraphics[height=4cm]{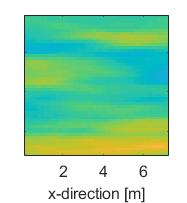}
			\caption{}
			\label{fig:res_lin2}
		\end{subfigure}
		\begin{subfigure}{.21\textwidth}
			\centering
		\includegraphics[height=4cm]{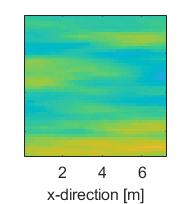}
			\caption{}
			\label{fig:res_lin3}
		\end{subfigure}
		\begin{subfigure}{.23\textwidth}
			\centering
		\includegraphics[height=4cm]{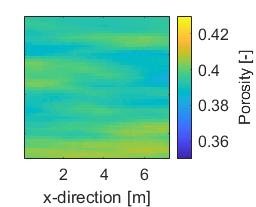}
			\caption{}
			\label{fig:res_lin1u}
		\end{subfigure}
		\end{flushleft}
		\begin{subfigure}{.32\textwidth}
			\centering
		\includegraphics[width=1\linewidth]{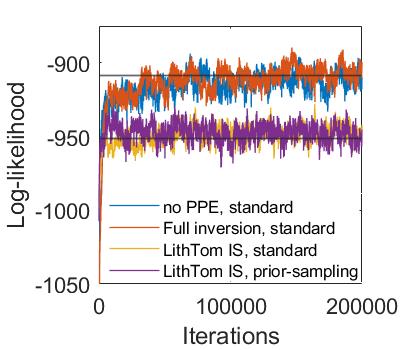}
			\caption{}
			\label{fig:res_lin7}
		\end{subfigure}	
		\begin{subfigure}{.32\textwidth}
			\centering
		\includegraphics[width=1\linewidth]{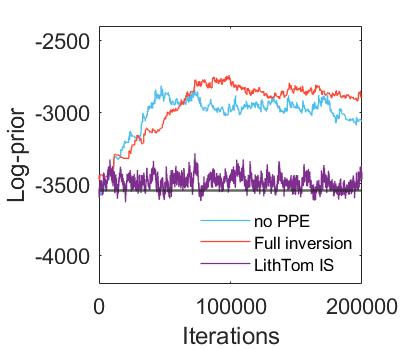}
			\caption{}
			\label{fig:res_lin8b}
		\end{subfigure}	\begin{subfigure}{.32\textwidth}
			\centering
		\includegraphics[width=1\linewidth]{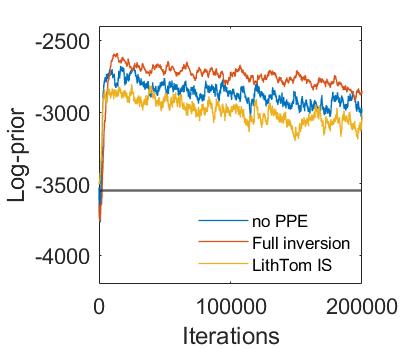}
			\caption{}
			\label{fig:res_lin8a}
		\end{subfigure}	
		\caption{Estimated posterior means of the porosity field $\por$ obtained for the linear test example with standard DREAM$_{(ZS)}$ proposals and (a) the algorithm ignoring PPEs, (b) the full inversion, (c) the LithTom IS method and with prior-sampling DREAM$_{(ZS)}$ proposals and (d) the LithTom IS method. (e) Corresponding log-likelihood values, black lines represent the values of $p(\data | \por, \latsmall)$ and $p(\data | \por)$ for the true porosity field $\por$ (and the true $\lat$ in the former). (f) Logarithmically transformed prior probabilities for the posterior samples obtained with prior-sampling DREAM$_{(ZS)}$ proposals and (g) standard DREAM$_{(ZS)}$ proposals; the black lines depict the prior probability of the true porosity field.}
		\label{fig:RESlin}		
	\end{figure}

Figure~\ref{fig:RESlin} presents the estimated posterior means of the porosity field obtained when applying the no PPE (Fig. \ref{fig:res_lin1}), the full inversion (Fig. \ref{fig:res_lin2}) and the LithTom IS (Fig. \ref{fig:res_lin3}) with standard DREAM$_{(ZS)}$ proposals, as well as for LithTom IS with prior-sampling DREAM$_{(ZS)}$ proposals (Fig. \ref{fig:res_lin1u}). These are the cases for which we reached convergence of the chains. The porosity field obtained with the inversion ignoring PPEs has, as expected (\citeauthor{brunetti} \citeyear{brunetti}), a higher variance. Visually, all other estimates are very similar in terms of structure and magnitude with respect to the analytical posterior mean in Figure \ref{fig:truepost1}. The estimated posterior mean of LithTom IS with the prior-sampling DREAM$_{(ZS)}$ proposals has a slightly lower variance than for standard DREAM$_{(ZS)}$ proposals. The ARs (Table \ref{tab:res1}) for standard DREAM$_{(ZS)}$ proposals are the highest for LithTom IS, while the method ignoring PPEs and full inversion have lower ARs.  Classical lithological tomography without importance sampling leads to an AR of less than 0.1~$\%$ such that, in practice, it unfeasible to reach convergence. Applying the CPM method without IS for $N$=50 and $\rho$=0.95 also results in an only slightly larger AR (roughly 0.2 $\%$), thereby, highlighting the need for importance sampling for the considered problem. Since less than 5 $\%$ of the parameters converged after 200'000 iterations, we renounce from showing further results for the CPM and PM method without IS. The method ignoring PPEs and the full inversion using prior-sampling DREAM$_{(ZS)}$ proposals suffer from very low ARs and did not reach convergence after 200'000 iterations. Table \ref{tab:res1} shows the number of iterations needed for the 99$^{th}$ percentile of the parameters' $\hat{R}$-statics to be below 1.2. It also shows the IACTs of the cell in the very middle of the porosity field for all inversion approaches reaching convergence within 200'000 MCMC iterations. We observe that the iterations needed for convergence and the IACT of the LithTom IS method with prior-sampling DREAM$_{(ZS)}$ proposals are the lowest. \\

Figure \ref{fig:res_lin7} shows the evolving log-likelihood values. When ignoring PPEs or performing the full inversion, the chains converge to much higher log-likelihoods than for the LithTom IS method. This is expected as they rely on the likelihood $p(\data | \por, \latsmall)$ (where $\lat = \mathcal{F}(\por) + \ppe$, with $\ppe = 0$ for the algorithm ignoring PPEs), while LithTom IS estimates $p(\data | \por) = \int p(\data | \por ,\latsmall) p(\latsmall | \por) d \latsmall$. This example highlights that LithTom IS broadens the likelihood function. Figures \ref{fig:res_lin8b} and \ref{fig:res_lin8a} show the prior probabilities (logarithmically transformed) for the posterior samples obtained with the three different inversion approaches using the two alternative proposal schemes. We observe that the LithTom IS method using prior-sampling DREAM$_{(ZS)}$ proposals (Fig. \ref{fig:res_lin8b}) is the only approach for which the prior probability of the true porosity field is sampled. All other methods and proposal scheme combinations sample porosity fields with higher prior probabilities than the true field (black solid line). Practically speaking, this implies for these cases that none of the posterior samples are close to the true model. Furthermore, the corresponding prior probabilities show a trend of slowly decreasing values raising doubts about the ergodicity of the MCMC chains. \\

\begin{table}[h]
\caption{\label{tab: table-name} Overview of the results obtained for the linear test example with the different inversion approaches and proposal mechanisms: The acceptance rates (\textbf{AR}), convergence (\textbf{Conv}) showing the number of iterations needed for the 99$^{th}$ percentile of the parameters' $\hat{R}$-statics to be below 1.2 (or the percentage of parameters with a $\hat{R}$-statistics below 1.2 if the the inversion did not converge), the mean KL-divergence (\textbf{KL-div}) and the integrated autocorrelation time (\textbf{IACT}) for the cell in the very middle of the porosity field $\por$.}
\begin{center}
\def\arraystretch{1.5}
\begin{tabular}{c|c|c|c|c|c|c}
	\textbf{Method} & \textbf{Proposal} & \textbf{Parameter} & \textbf{AR }&\textbf{Conv} &  \textbf{KL-div}  &  \textbf{IACT} \\
	\hline \hline
	No PPE & Standard & - & 10 $\nearrow$ 20 $\%$ & 104'000 & 1.957 & 3'850 \\
	\hline
	LithTom  & Standard & $N=1, \rho=0$ & < 0.1 $\%$ & - , 0 $\%$ & - & -\\
	\hline
	CPM no IS & Standard & $N=10, \rho=0.95$ & 0.1 $\%$ & - , 3 $\%$ & - & - \\
	 & Standard & $N=50, \rho=0.95$  & 0.2 $\%$ & - , 4 $\%$ & \\
	\hline
	Full inversion & Standard & - & 10 $\nearrow$ 20 $\%$ & 150'000 & 0.354 & 6'900 \\
	\hline
	LithTom IS & Standard & $N=1, \rho=0$ & 20 $\nearrow$ 30 $\%$ & 78'000 & 0.063 & 2'750 \\
	\hline \hline
	no PPE & Prior-sampling & - & 1 - 2 $\%$ &  - , 35 $\%$ & - & -  \\
	\hline
	Full inversion & Prior-sampling & - & 1 - 2 $\%$ & - , 14 $\%$ & - & -  \\
	\hline
	LithTom IS & Prior-sampling & $N=1, \rho=0$ & 13 $\%$ & 76'000 & 0.003 & 1'700
\end{tabular}
\end{center}
\label{tab:res1}
\end{table}

\begin{figure}[h]
\begin{flushright}
		\begin{subfigure}{.26\textwidth}
			\centering
			\includegraphics[height=4cm]{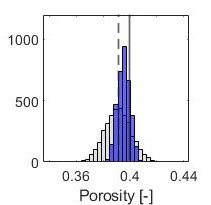}
			\caption{}
			\label{fig:res_lin2a}
		\end{subfigure}
		\begin{subfigure}{.23\textwidth}
			\centering
			\includegraphics[height=4cm]{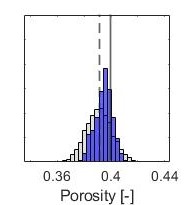}
			\caption{}
			\label{fig:res_lin2b}
		\end{subfigure}
		\begin{subfigure}{.23\textwidth}
			\centering
			\includegraphics[height=4cm]{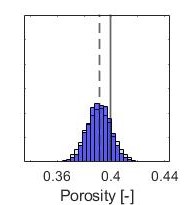}
			\caption{}
			\label{fig:res_lin2c}
		\end{subfigure}
	\begin{subfigure}{.23\textwidth}
		\centering
		\includegraphics[height=4cm]{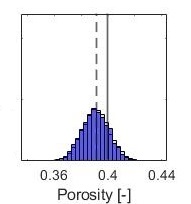}
		\caption{}
		\label{fig:res_lin2e}
	\end{subfigure}
\end{flushright}
		\caption{Histograms comparing samples from the analytical posterior PDF $p(\por|\data)$ (light grey) for the linear test example and samples from the respective inversion method (blue), the solid line depicts the true value of the porosity in the very middle of the model domain and the dashed line indicates the analytical posterior mean (a) no PPE and standard DREAM$_{(ZS)}$ proposals, (b) full inversion and standard DREAM$_{(ZS)}$ proposals, (c) LithTom IS and standard DREAM$_{(ZS)}$ proposals and (d) LithTom IS and prior-sampling DREAM$_{(ZS)}$ proposals.}
		\label{fig:RESlin2}				
\end{figure}

\begin{figure}[h]
\begin{flushright}
		\begin{subfigure}{.225\textwidth}
			\centering
			\includegraphics[height=4cm]{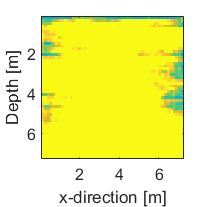}
			\caption{}
			\label{fig:res_lin3a}
		\end{subfigure}
		\begin{subfigure}{.21\textwidth}
			\centering
			\includegraphics[height=4cm]{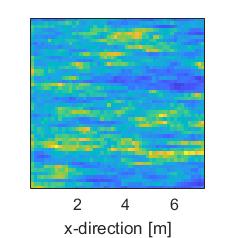}
			\caption{}
			\label{fig:res_lin3b}
		\end{subfigure}
		\begin{subfigure}{.21\textwidth}
			\centering
			\includegraphics[height=4cm]{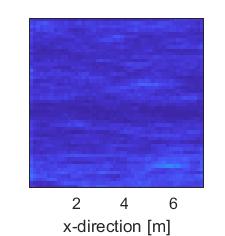}
			\caption{}
			\label{fig:res_lin3c}
		\end{subfigure}
	\begin{subfigure}{.31\textwidth}
		\centering
		\includegraphics[height=4.25cm]{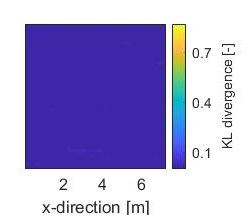}
		\caption{}
		\label{fig:res_lin3e}
	\end{subfigure}
\end{flushright}
		\caption{KL-divergences with respect to the analytical posterior PDF $p(\por|\data)$ for the linear test example (a) no PPE and standard DREAM$_{(ZS)}$ proposals, (b) full inversion and standard DREAM$_{(ZS)}$ proposals, (c) LithTom IS and standard DREAM$_{(ZS)}$ proposals and (d) LithTom IS and prior-sampling DREAM$_{(ZS)}$ proposals.}
\label{fig:RESlin3}				
\end{figure}

To compare the posterior PDFs with the analytical solution, we consider first histograms for an exemplary position in the porosity field and the KL-divergences of the whole field. We only show the results of the method and proposal-scheme combinations that converged within the considered 200'000 iterations. The histograms are depicted in Figure \ref{fig:RESlin2} with samples from the analytical posterior PDF (light grey) and samples from the respective inversion method (blue) for the pixel in the very middle of the model domain. The corresponding KL-divergences for all pixels are shown in Figure \ref{fig:RESlin3}. The histogram and the KL-divergences of the method ignoring PPEs (with standard DREAM$_{(ZS)}$; Figures \ref{fig:res_lin2a} and \ref{fig:res_lin3a}) indicate that the approach suffers from biased estimates and an underestimation of the posterior variance. The posterior samples obtained with the full inversion method (with standard DREAM$_{(ZS)}$ proposals; Figures \ref{fig:res_lin2b} and \ref{fig:res_lin3b}) better represent the analytical posterior PDF, but there is still a significant underestimation of the posterior variance. The histogram obtained with the LithTom IS approach using standard DREAM$_{(ZS)}$ proposals (Figure \ref{fig:res_lin2c}) is very similar to the one of the analytical posterior. The corresponding six-fold decreases of the KL-divergence (Figure \ref{fig:res_lin3c}) compared with full inversion confirm the significant improvements of the exploration capabilities of this approach. An even better representation of the analytical posterior was obtained with the LithTom IS approach when using prior-sampling DREAM$_{(ZS)}$ proposals. This is indicated by the histogram in Figure \ref{fig:res_lin2e} and by a further two-fold decrease of the KL-divergence in Figure \ref{fig:res_lin3e}. An overview of the mean KL-divergences is given in Table \ref{tab:res1}. \\

This linear example has been used to show that importance sampling and prior-preserving proposal schemes are essential to obtain meaningful results in our considered high-dimensional setting. For this example, one can get accurate results using LithTom IS alone. The next section dealing with the non-linear case will serve to demonstrate the benefits of the CPM method in non-linear settings.   \\

\subsection{Non-linear physics}  
\label{nonlinear}
We now consider a non-linear test case in which the 625 arrival times are generated with the eikonal 2D traveltime solver \textit{time2D} of \citet{time2D} such that,	
    \begin{equation}
    \data = \mathcal{G}_{e}(\latsmall) + \datanoise. 
    \end{equation}
Given the non-linear physics, the likelihood function $p(\data|\por)$ is intractable and there is no analytical expression for the posterior PDF $p(\por | \data)$ to compare with. The same applies for the PDF $p(\latsmall|\data, \por)$ that we previously used for the importance sampling of the latent variable~$\lat$. Hence, as importance sampling distribution we rely on the approximation of the PDF $p(\latsmall|\data, \por)$ introduced in Section \ref{S.IS}. For $\latsmall_{lin} = \mathcal{F}(\por_{lin}) + \ppe_{lin} = \frac{1}{c} \left( \sqrt{\ks} + (\sqrt{\kw}-\sqrt{\ks}) \por_{lin} \right) + \ppe_{lin}$, we use the last state of the porosity field for $\por_{lin}$ and the previous importance sampling mean $\muIS$ for $\ppe_{lin}$. To decrease computational resources, we only update the linearization every 100 MCMC iterations. Since the expression is approximate, we further inflate the importance sampling covariance matrix $\sigmaIS$ by multiplying $\boldsymbol{\Sigma_{\dataRV}}$ with a factor. After initial testing, we found that $1.2$ yielded the best performance. \\
	
\begin{figure}[h]
	\begin{subfigure}{.48\textwidth}
			\centering
			\includegraphics[width=1\linewidth]{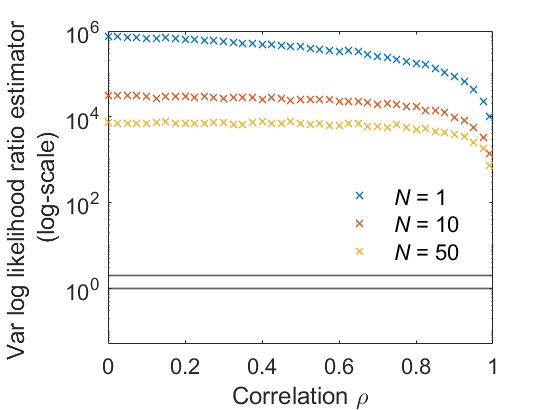}
			\caption{}
			\label{fig:logLH2}	
		\end{subfigure}
		\begin{subfigure}{.48\textwidth}
			\centering
			\includegraphics[width=1\linewidth]{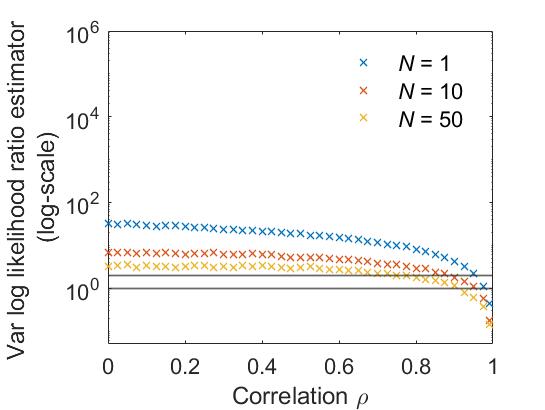}
			\caption{}
			\label{fig:logLH1}	
		\end{subfigure}
		\caption{Variance of the log likelihood ratio estimator $R=\log \left(\hat{p}_N^{(j)}(\data|\por) \right)  - \log \left( \hat{p}_N^{(j-1)}(\data|\por) \right)$ for the non-linear test example and $\por$ fixed at a region with high posterior probability mass as a function of $\rho$ (used to correlate the latent variables $\lat^{(j)}$ and $\lat^{(j-1)}$ as in Equation (\ref{corr})) for $N=1$, $N=10$ and $N=50$ samples of the latent variable $\lat$; the realizations of the latent variable are drawn (a) from the prior $p(\latsmall | \por)$ and (b) with importance sampling. The black lines delimit the range between $1.0$ and $2.0$ recommended by \citet{CPM}.}
		\label{fig:logLH}		
	\end{figure}

Figure \ref{fig:logLH} depicts the dependence of the variance of the log-likelihood ratio estimator $R$ (Eq.~(\ref{loglhesti})) on the correlation parameter $\rho$ for $N=1$, $N=10$ and $N=50$ samples of the latent variable $\lat$ (with $\por$ being fixed at a region with high posterior probability mass). Figure \ref{fig:logLH2} depicts estimates when drawing the realizations of the latent variable proportionally to its prior distribution $p(\latsmall|\por)$ and Figure \ref{fig:logLH1} for the case where the latent variable is sampled with importance sampling. The two plots highlight three fundamental aspects of the CPM method in our geophysical setting. First, it is crucial to use a well chosen importance sampling for the latent variable draws, since for a correlation of, say, $\rho=0$, the variance of the log likelihood ratio estimator can be reduced from values between $10'000$ and $1'000'000$ (using sampling from prior) to values between $3$ and $31$ (using importance sampling). Second, increasing the number of draws of latent variables ($N$) decreases the variance of the log-likelihood ratio estimator further and, third, this is also achieved by increasing the amount of correlation ($\rho$) used for two subsequent draws of latent variables. The variance for $\rho=1$ is equal to zero for all parameter settings (as we use the same values for $\lat^{(j-1)}$ and $\lat^{(j)}$). Without importance sampling, we could still obtain a variance of the log-likelihood ratio estimator between 1 and 2 as recommended by \citet{CPM}, but with the need of a very high $N$ or a $\rho$ very close to 1. In practice, this would either result in excessively high computational costs or slow mixing in the draws of the latent variables. \\

Due to the high variances displayed in Figure \ref{fig:logLH2} and since the pseudo-marginal approaches without importance sampling have already proven to be highly inefficient in the linear case (Table \ref{tab:res1}), we now restrict ourselves only to CPM implementations involving IS. In stark contrast to the linear case, the LithTom IS approach ($N=1, \rho=0$) leads to a highly inefficient algorithm, as the variance of $R$ around 30 is much higher than the upper recommended threshold of $2.0$. For the CPM method, we set the number of samples to $10$ and the correlation to $\rho=0.95$ as this values leads to a variance of the log likelihood ratio estimator in-between $1.0$ and $2.0$. The autocorrelation of one cell of the latent variable field is given by $Corr(X_1, X_{1+l}) = \rho^{l}$ for lag $l$ with the correlation mechanism of Equation (\ref{corr}), such that for $\rho=0.95$ roughly $100$ (accepted) iterations are needed to draw an independent realization of the latent variable. In practice, the decorrelation will be slower as we only move on with accepted proposals (Section \ref{sec_cpm}) .\\

\begin{figure}[h]	
\begin{flushleft}
		\begin{subfigure}{.225\textwidth}
			\centering
			\includegraphics[height=4cm]{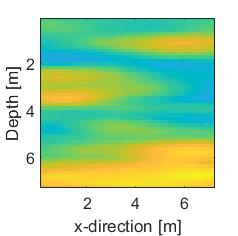}
			\caption{}
			\label{fig:res_nlin1}
		\end{subfigure}
		\begin{subfigure}{.21\textwidth}
			\centering	\includegraphics[height=4cm]{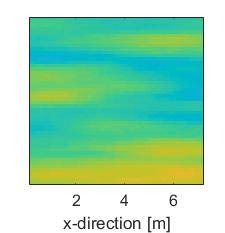}
			\caption{}
			\label{fig:res_nlin2}
		\end{subfigure}
		\begin{subfigure}{.21\textwidth}
			\centering
	\includegraphics[height=4cm]{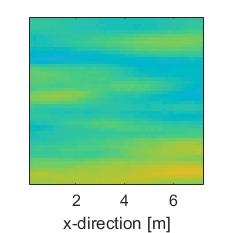}
			\caption{}
			\label{fig:res_nlin3}
		\end{subfigure}
		\begin{subfigure}{.23\textwidth}
			\centering
	\includegraphics[height=4cm]{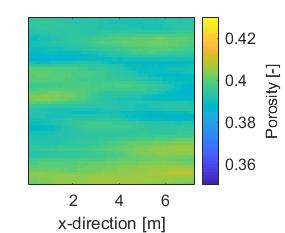}
			\caption{}
	\label{fig:res_nlin1u}
		\end{subfigure}
		\end{flushleft}
		\begin{subfigure}{.32\textwidth}
			\centering
	\includegraphics[height=5cm]{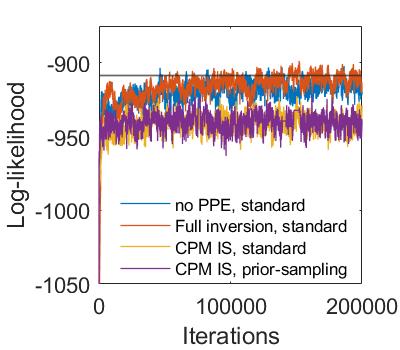}
			\caption{}
			\label{fig:res_nlin5}
		\end{subfigure}
		\begin{subfigure}{.32\textwidth}
			\centering
			\includegraphics[height=5cm]{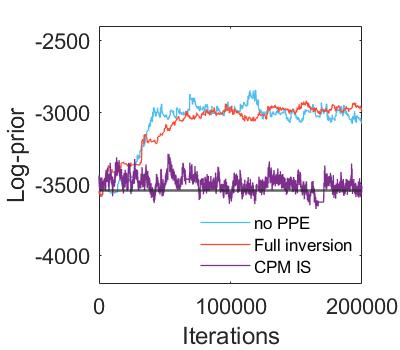}
			\caption{}
			\label{fig:res_nlin6}
		\end{subfigure}
		\begin{subfigure}{.32\textwidth}
			\centering
			\includegraphics[height=5cm]{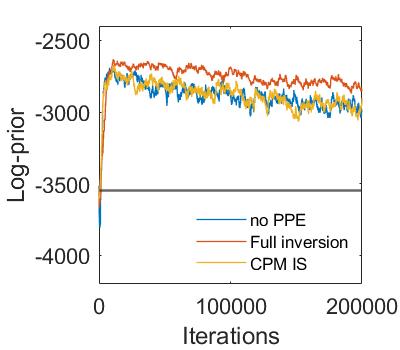}
			\caption{}
			\label{fig:res_nlin7}
		\end{subfigure}	
		\caption{Estimates of the posterior means of the porosity field $\por$ for the non-linear test example resulting with standard DREAM$_{(ZS)}$ proposals and (a) the algorithm ignoring PPEs, (b) the full inversion, (c) the CPM IS ($N=10$, $\rho = 0.95$) method. Results for prior-sampling DREAM$_{(ZS)}$ proposals and (d) the CPM IS ($N=10$, $\rho = 0.95$) method. (e) Log-likelihood functions, black line represents the value of $p(\data | \por, \latsmall)$ for the true porosity and latent variable field. (f) Prior probabilities (logarithmically transformed) of the posterior samples obtained with prior-sampling DREAM$_{(ZS)}$ proposals and (g) standard DREAM$_{(ZS)}$ proposals; the black lines depict the corresponding value for the true porosity field.}
		\label{fig:RESnlin}				
	\end{figure}

The results for both DREAM$_{(ZS)}$ proposal schemes are shown in Figure~\ref{fig:RESnlin} and Table~\ref{tab:res3}. For the estimates of the posterior mean of the porosity field (Fig. \ref{fig:res_nlin1}-\ref{fig:res_nlin1u}), we observe similar results as in the linear case: Using prior-sampling DREAM$_{(ZS)}$ proposals results in a porosity field estimate with lower variance and using the method ignoring PPEs (Fig. \ref{fig:res_nlin1} for standard DREAM$_{(ZS)}$ proposals) leads to higher variance. The highest acceptance rate is obtained with applying the CPM IS method using standard DREAM$_{(ZS)}$ proposals (Table ~\ref{tab:res3}) and the acceptance rates for prior-sampling DREAM$_{(ZS)}$ proposals are lower. The LithTom approach with IS has an AR of less than 1 $\%$ and would, therefore, require far more than 200'000 iterations to converge. Trace plots of the evolving log-likelihood values are shown in Figure \ref{fig:res_nlin5}. As expected and in agreement with the linear test case (Fig. \ref{fig:res_lin7}), the methods converge to different values. As in  the linear case, we find that CPM IS with prior-sampling DREAM$_{(ZS)}$ proposals is the only case providing posterior samples that match the prior probability of the true porosity field (Fig.~\ref{fig:res_nlin6} and \ref{fig:res_nlin7}). \\

\begin{table}[h]
\caption{\label{tab: table-name} Summary of the results obtained for the non-linear test example with the various inversion approaches and the two proposal mechanisms: The acceptance rates (\textbf{AR}), the convergence (\textbf{Conv}) showing the number of iterations needed for the 99$^{th}$ percentile of the parameter's $\hat{R}$-statics to be below 1.2 (or the percentage of parameters with a $\hat{R}$-statistics below 1.2 if the the algorithm did not converge), the percentage of pixels in which the true porosity value lies within the range of posterior samples ($\textbf{$\por_{true}$}$), the mean logarithmic score (\textbf{logS}), posterior standard deviation (\textbf{Post SD}) and the integrated autocorrelation time (\textbf{IACT}) for the cell in the very middle of the porosity field $\por$. The CPM IS method was evaluated with the parameter choice of $N=10$ and $\rho=0.95$. }
\begin{center}
\def\arraystretch{1.5}
\begin{tabular}{c|c|c|c|c|c|c|c}
	\textbf{Method} & \textbf{Proposal}  & \textbf{AR }&\textbf{Conv} & \textbf{$\por_{true}$} & \textbf{logS} & \textbf{Post SD} &  \textbf{IACT}  \\
	\hline \hline
	No PPE & Standard  &  11 $\nearrow$ 24 $\%$ & 92'000  & 87.2 $\%$ & 3.36 & $5.4 \times 10^{-3}$ & 3'800 \\
	\hline 
	Full inversion & Standard  & 10 $\nearrow$ 23 $\%$ & 144'000  & 97.1 $\%$ & 1.99 & $6.7 \times 10^{-3}$ & 5'150\\
	\hline
	LithTom IS & Standard  & < 1 $\%$ & - , 43 $\%$  & - & - & - & - \\
	\hline
	CPM IS & Standard & 12 $\nearrow$ 24 $\%$ & 90'000  & 99.6 $\%$ & 1.56 & $8.3 \times 10^{-3}$ & 3'250 \\
	 \hline \hline 
	 No PPE & Prior-samp &   1 - 2 $\%$ & - , 29 $\%$ & - & - & - & - \\
	\hline 
	Full inversion & Prior-samp  & 1 - 2 $\%$ & - , 13 $\%$ & - & - & - & - \\
	\hline
	CPM IS & Prior-samp  & 11 $\%$ & 96'000  & 100.00 $\%$ & 1.34 & $10.4 \times 10^{-3}$ & 3'300 \\
\end{tabular}
\end{center}
\label{tab:res3}
\end{table}
	
\begin{figure}[h]
\begin{flushleft}		
\begin{subfigure}{.23\textwidth}
			\centering
			\includegraphics[height=4cm]{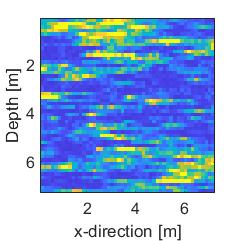}
			\caption{}
			\label{fig:PIT1}
		\end{subfigure}
	\begin{subfigure}{.23\textwidth}
		\centering
		\includegraphics[height=4cm]{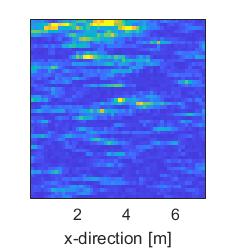}
		\caption{}
		\label{fig:PIT2}
	\end{subfigure}
	\begin{subfigure}{.21\textwidth}
		\centering
	\includegraphics[height=4cm]{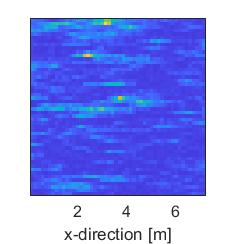}
		\caption{}
		\label{fig:PIT3}
	\end{subfigure}
		\begin{subfigure}{.29\textwidth}
	\centering
	\includegraphics[height=4cm]{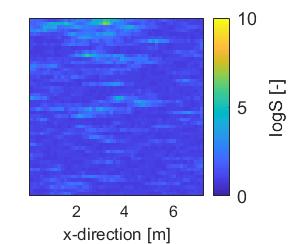}
		\caption{}
		\label{fig:PIT5}
	\end{subfigure}
	\end{flushleft}	
\caption{The logarithmic scores for the non-linear test case with (a) no PPE and standard DREAM$_{(ZS)}$ proposals, (b) full inversion and standard DREAM$_{(ZS)}$ proposals, (c) CPM IS and standard DREAM$_{(ZS)}$ proposals and (d) CPM IS and prior-sampling DREAM$_{(ZS)}$ proposals.}
\label{fig:pit}				
\end{figure}

Figure \ref{fig:pit} depicts the logarithmic scores (see Section \ref{assessment}) comparing the true porosity values with the inferred posterior PDFs for all 2500 grid cells. We observe that the method ignoring PPEs (with standard DREAM$_{(ZS)}$ proposals, Fig. \ref{fig:PIT1}) has the highest scores (indicating the lowest accuracy). The values of the full inversion (with standard DREAM$_{(ZS)}$ proposals, Fig. \ref{fig:PIT2}) are lower, but still high. The CPM IS method with standard DREAM$_{(ZS)}$ proposals (Figs. \ref{fig:PIT3}) leads to reduced logarithmic scores that are further improved when this method is combined with prior-sampling DREAM$_{(ZS)}$ proposals (Figs. \ref{fig:PIT5}). The mean values of the logarithmic scores and other performance metrics are shown in Table~\ref{tab:res3}. We find that the method that ignores PPEs fails to sample a range of values including the true porosity value in more than 10$\%$ of the pixels and has a mean estimated posterior standard deviation that is up to 50 $\%$ smaller than the other methods. The CPM IS method generates posterior samples with ranges that include, in more than 99 $\%$ of the pixels, the true porosity value with the percentages obtained using prior-sampling DREAM$_{(ZS)}$ proposals being even higher. Finally, the full inversion does not sample the true porosity value in almost 3$\%$ of the pixels and has a reduced mean estimated posterior standard deviation by up to 40 $\%$ compared to the CPM IS method. We also note that the IACT of the CPM methods are the lowest (Table \ref{tab:res3}). \\

\section{Discussion}
\label{4}
This study showed clearly that the correlated pseudo-marginal (CPM) method, which accounts for petrophysical prediction uncertainty within the estimate of the likelihood function~$p(\data|\por)$, combined with importance sampling (IS) and prior-sampling MCMC proposals leads to a broader exploration of the target posterior $p(\por|\data)$ than the other presented combinations of inversion methods and proposal schemes. The CPM method is an exact and general method, but it needs in the considered high-dimensional setting an efficient importance sampling and prior-sampling proposals to work well even for the case of linear physics. \\

In the linear setting (with available analytical solutions for the PDFs), the CPM method using importance sampling performs well using only one uncorrelated sample of the PPE (LithTom IS). In absence of importance sampling, even a high number of samples $N$ and correlation $\rho$ could not prevent the algorithm from being highly inefficient (Table \ref{tab:res1}). We find that the exploration of the posterior PDF is much improved when using the LithTom IS approach compared with full inversion (Fig. \ref{fig:RESlin2} and Fig. \ref{fig:RESlin3}). Although the $\hat{R}$-statistic of \citet{rstat} suggests that the full inversion algorithm (using standard DREAM$_{(ZS)}$ proposals) has converged, we demonstrate a significant underestimation of the posterior standard deviation and posterior samples with far too high prior probabilities compared with the true model (Fig. \ref{fig:res_lin8b} and \ref{fig:res_lin8a}). Indeed, the full inversion's high acceptance rate (for standard DREAM$_{(ZS)}$ proposals) may be mainly a consequence of local exploration combined with an adaptive MCMC expanding its archive. This (1) points out that  Gelman-Rubin's $\hat{R}$-statistics and the acceptance rate are insufficient metrics to assess the performance of an adaptive MCMC algorithm such as DREAM$_{(ZS)}$ and (2) highlights issues with over-fitting when using adaptive MCMC. Indeed, \citet{robert2018} warn against using adaptive MCMC methods without due caution as adaptations to the proposal scheme can lead to algorithms relying too much on previous iterations, thereby, excluding parts of the parameter space that have not yet been explored. \\

The need for a well-chosen importance sampling distribution is also demonstrated for the non-linear setting by analysing the variances of the log-likelihood ratio estimator (Fig.~\ref{fig:logLH}). This analysis also confirmed the strong influence of $N$ and $\rho$. Since the importance sampling distribution is no longer exact in the non-linear test case, the number of samples $N$ and the correlation $\rho$ need to be increased. Consequently, the CPM IS method performs better (in terms of computational cost) than the PM IS method as fewer samples have to be used. For the non-linear test case, we conclude that the exploration of the posterior with the CPM IS method (especially when combined with prior-sampling DREAM$_{(ZS)}$ proposals) is better than the full inversion by observing that (1) the range of the posterior samples includes more often the true porosity value while (2) the logarithmic score is lower and (3) the mean estimated posterior standard deviation is higher (Table \ref{tab:res3}).  \\

 We recommend to work in the full parameter space whenever possible such that any distortions in the posterior estimations due to model reductions can be avoided. The presented adaptive prior-preserving proposal scheme (prior-sampling DREAM$_{(ZS)}$ proposal) is developed in the spirit of the extended Metropolis algorithm of \citet{mosegaard} and the pCN proposal of \citet{cotter2013}. It is a simple correction of the standard DREAM$_{(ZS)}$ proposal that (1) makes the algorithm robust to the choice of the discretization of the target field and (2) maintains its capabilities to sample efficiently in complex high-dimensional parameter spaces. We find that the prior-sampling DREAM$_{(ZS)}$ proposals lead to an enhanced exploration of the posterior PDF and a stable AR (Tables \ref{tab:res1} and \ref{tab:res3}). Indeed, the CPM IS approach with prior-sampling proposals is the only one generating samples with a prior probability comparable to the one of the true porosity field (Figs. \ref{fig:RESlin} and \ref{fig:RESnlin}). Due to dependencies between latent and target variables, the full inversion with prior-sampling DREAM$_{(ZS)}$ proposals suffers from a very low acceptance rate as the method does not allow for large proposal steps. This dependency is bypassed by the CPM IS, allowing larger steps for a given AR. In general, combinations of adaptive Metropolis and pCN-proposals are referred to as DIAM (dimension independent adaptive Metropolis) proposals and were introduced by \citet{chen2016}. Another way to increase the efficiency of the pCN proposal was proposed by \citet{rudolf} with the so-called generalized pCN-proposal (gpCN), in which the proposal scheme is tuned to have the same covariance as the target posterior distribution. \\

We emphasize that this study only considers synthetic data. We demonstrate that all but our method of choice (CPM IS with prior-sampling DREAM$_{(ZS)}$ proposals) have severe problems in exploring the full posterior distribution even in this well-specified setting. A field demonstration of CPM IS with prior-sampling DREAM$_{(ZS)}$ proposals is a natural next step. Furthermore, our entire study remains within Gaussian assumptions for the target field, petrophysical prediction uncertainty and observational noise. In the presented results, we deal only with weak non-linearity in our forward operator and assume the petrophysical relationship to be linear. In the future, it would be useful to consider test cases involving stronger non-linearity, be it through a higher variance of the slowness field or a non-linear petrophysical relationship. Stronger non-linearity would affect the accuracy of the first-order expansion used to derive the importance sampling distribution for the CPM method, implying that the approximations would become less accurate. This could lead to a decrease of efficiency that could be counter-acted by using larger $N$ or $\rho$. An important topic for future research would be to develop and assess importance sampling schemes that do not rely on Gaussian assumptions. Potential starting points could be efficient importance sampling by \citet{richard} or multiple importance sampling introduced by \citet{veach} and popularised by \citet{IS}. \\

In agreement with \citet{brunetti}, we find that ignoring petrophysical prediction uncertainty leads to biased estimates and too tight uncertainty bounds. While the need for a method accounting for PPEs grows with increasing integral scale of the target field (\citeauthor{brunetti} \citeyear{brunetti}), the ratio of the variances of the PPE, the target variable and the observational noise also influences the results. The need for a well-working importance sampling for CPM grows with increasing petrophysical prediction uncertainty and decreasing observational noise. At the same time, large petrophysical prediction uncertainty leads to a flattened likelihood function $p(\data|\por)$, thereby, decreasing the variance of the likelihood estimators (assuming a well-working importance sampling) and, therefore, enhancing the efficiency of the algorithm. Our present work focuses on petrophysical prediction uncertainty for a known covariance model, but it would be possible to expand this to an unknown covariance model, an uncertain petrophysical model or uncertain model parameters. 

\section{Conclusions}
\label{5}

We consider lithological tomography in which geophysical data are used to infer the posterior PDF of target (hydro)geological parameters. In such a latent variable model, the geophysical properties play the role of latent variables that are linked to the properties of interest through petrophysical relationships exhibiting significant scatter. Compared with the original formulation of lithological tomography that does not consider importance sampling, we make the approach more applicable to high dimensions (thousands of unknowns) and large data sets with high signal-to-noise ratios. To account for the intractable likelihood appearing in the Metropolis--Hastings algorithm in this setting, we explore the correlated pseudo-marginal (CPM) method using an importance sampling distribution and prior-sampling proposals. For the latter, we adapt the standard (adaptive) proposal scheme of DREAM$_{(ZS)}$ with a prior-sampling approach, leading to a further improvement in exploration compared with standard model proposals when dealing with high-dimensional problems. We find that our implementation of the CPM method outperforms standard lithological tomography and the full inversion approach, which parameterizes and infers the posterior petrophysical prediction uncertainty. For a linear test example, the mean KL-divergence with respect to the analytical posterior can be reduced by 99 $\%$ by our implementation of the CPM method (even without using correlations) compared with full inversion. In the case of non-linear physics, we reduce the mean logarithmic score with respect to the true porosity field by up to 33 $\%$ compared with the full inversion method. The CPM method is generally applicable and accurate, but it requires a well-working importance sampling distribution (presently based on Gaussian random field theory) to be efficient. Future work with the CPM method could consider field data applications, more non-linear physics and non-linear petrophysical relationships as well as relaxing the assumptions of Gaussian random fields. Furthermore, the method's use in coupled hydrogeophysical inversions involving hydrogeological flow and transport models would be of interest.

\section*{Acknowledgements}
This work was supported by the Swiss National Science Foundation (project number: \href{http://p3.snf.ch/project-184574}{184574}). We are grateful for the constructive comments offered by associate editor Juan Carlos Afonso, Andrea Zunino and an anonymous reviewer.  

\section*{Data Availability}
No new data were generated or analysed in support of this research.

\pagebreak
\appendix

\section{DREAM algorithms and prior-sampling proposals}
\label{appendix_dream}
To perform a high-dimensional inversion with the MH algorithm, one needs a well-working proposal scheme. To deal with this challenge, \citet{terbraak2006} introduced an adaptive random walk MH algorithm named Differential Evolution Markov chain (DE-MC). This method runs $C$ Markov chains in parallel, where at each iteration $j$, the $C$ different realizations of the model parameters define a population $\{\boldsymbol{Z}^{(j)}_c; c=1,2,...,C\}$, which is used to guide new model proposals. For chain $c$, two chains (denoted as $a$ and $b$) are drawn without replacement from the remaining set of chains. Then, the algorithm proposes a new state for the $c$-th chain with,	
\begin{equation}
		\boldsymbol{Z}^{(j)}_c = \boldsymbol{Z}^{(j-1)}_c + \gamma(\boldsymbol{Z}^{(j-1)}_a - \boldsymbol{Z}^{(j-1)}_b) + \zeta, \quad c \neq a \neq b
\end{equation}
where $\gamma$ denotes the jumping rate and $\zeta$ is a draw from $\mathcal{N}(0,s^2)$ with a small standard deviation~$s$ used to ensure that the resulting Markov chain is irreducible. By accepting or rejecting the resulting proposals with the MH-ratio of Equation~(\ref{MHr}), a Markov chain with the posterior PDF as its stationary distribution is obtained (Proof in \citeauthor{DREAMproof} \citeyear{DREAMproof}). This leads to an algorithm which is automatically adapting the scale and the orientation of the proposal density along the way to the stationary distribution, allowing it to provide efficient sampling on complex, high-dimensional, and multi-modal target distributions. Based on the DE-MC, \citet{DREAM1} introduced the adaptive multi-chain MCMC algorithm called DREAM (DiffeRential Evolution Adaptive Metropolis). It enhances the efficiency of DE-MC by applying subspace sampling (only randomly selected dimensions of the model parameter are updated) and outlier chain correction.  An excellent overview of the theory and application of the DREAM algorithm is given by \citet{DREAM}. For our case study, we use the extended version DREAM$_{(ZS)}$ introduced by \citet{dream_zs}, as its proposal scheme using an archive of past states leads to further improved convergence and posterior exploration. \\

To adapt extended Metropolis to DREAM$_{(ZS)}$, we rely on a transformation of the variables to the Uniform space. In our case study with Gaussian target variable $\boldsymbol{Z}^{(j)}_c~=~(Z^{(j)}_{c;1}, Z^{(j)}_{c;2},..., Z^{(j)}_{c;D^2})$ sampled in chain $c$ and iteration $j$, we define $U^{(j)}_{c;i}~=~\Phi(Z^{(j)}_{c;i})$, with $\Phi(\cdot)$ being the standard-normal cumulative distribution function (CDF), and apply the proposal mechanism of DREAM$_{(ZS)}$ on this transform. Assuming that $Z^{(j)}_{c;i}$ has a standard-normal distribution, $U^{(j)}_{c;i}$ will be distributed uniformly on $[0,1]$. The proposal scheme of DREAM$_{(ZS)}$ with so-called fold boundary handling (i.e., periodic boundary conditions) ensures that the new state $U^{(j+1)}_{c;i}$ is a sample from the Uniform distribution as well. With the subsequent transformation back to the standard normal, $Z^{(j+1)}_{c;i} = \Phi^{-1}(U^{(j+1)}_{c;i})$, we hence force the algorithm to use a proposal scheme that samples from the prior PDF.

\section{Analytical posterior PDF and importance density for linear physics}
\label{appendix2}

Assuming linear physics and petrophysics, it is possible to derive an analytical expression for the posterior PDF~$p(\por | \data)$ of the porosity (or other variable of interest). We consider here both relationships being linear without intercept ($\mathcal{G}(\lat) = \boldsymbol{J}_s \lat$ and $\mathcal{F}(\por) = \boldsymbol{J}_p \por$), however, an intercept (as the one used for $\mathcal{F}(\por)$ in our test case; Section \ref{data}) is easily included. For the 2D grid of the porosity $\por$ and the latent variable $\lat$, we use the following prior PDFs: 
\begin{equation}
	\label{prior}
	p(\por) = \varphi_{D^2}(\por; \pormu, \porsigma), \quad
	p(\latsmall | \por) = \varphi_{L}(\latsmall; \boldsymbol{J}_p \por, \ppesigma).
\end{equation}
To derive the (in this case) tractable likelihood $p(\data | \por)$, we use a standard result about marginal and conditional Gaussians (\citeauthor{bishop} \citeyear{bishop}): 

\vspace{1cm}
\begin{lemma}
Marginal and Conditional Gaussians

	Assume a  marginal Gaussian distribution for $\mathbf{X} \in \mathbb{R}^L$ and a conditional Gaussian distribution for $\mathbf{Y} \in \mathbb{R}^T$ given~$\mathbf{X}$ in the form
	\begin{align*}
		&p(\mathbf{x}) = \varphi_T(\mathbf{x};\boldsymbol{\mu}, \mathbf{\Lambda^{-1}}), \\
		&p(\mathbf{y}|\mathbf{x}) = \varphi_T(\mathbf{y};\mathbf{Ax+b}, \mathbf{L^{-1}}),
	\end{align*}
	with $\varphi_{T}(\cdot; \boldsymbol{\mu}, \boldsymbol{K})$ denoting the PDF of the $T$-variate Normal distribution with mean $\boldsymbol{\mu}$ and covariance matrix $\boldsymbol{K}$. Then, the marginal distribution of $\mathbf{Y}$ and the conditional distribution of $\mathbf{X}$ given $\mathbf{Y}$ are given by
	\begin{align}
		\label{marg}
		&p(\mathbf{y}) = \varphi_T(\mathbf{y};\mathbf{A \boldsymbol{\mu} + b}, \mathbf{L^{-1} + A\Lambda^{-1}A^T}) \\
		\label{cond}
		&p(\mathbf{x}|\mathbf{y}) = \varphi_L(\mathbf{x};\mathbf{\Sigma\left(A^TL(y-b) + \Lambda \boldsymbol{\mu} \right)}, \mathbf{\Sigma})
	\end{align}
	where 
	\begin{equation*}
		\mathbf{\Sigma = (\Lambda + A^TLA)^{-1}}.
	\end{equation*}
\end{lemma}
\vspace{1cm}

Using the prior on the latent variable $\lat$ and the Gaussian likelihood  $p(\data |\latsmall, \por) = \varphi_{625}(\data; \boldsymbol{J}_s \latsmall, \boldsymbol{\Sigma_{\dataRV}})$, we get with Equation (\ref{marg}), 
\begin{equation}
	p(\data | \por) = \varphi_{T} (\data; \boldsymbol{J}_s \boldsymbol{J}_p \por, \boldsymbol{\Sigma_{\dataRV}} + \boldsymbol{J}_s \ppesigma \boldsymbol{J}_s^T ).
\end{equation}
Subsequently, the analytical form of the posterior $p(\por | \data)$ is derived with Equation (\ref{cond}), the prior on porosity and the expression of the likelihood $p(\data | \por)$ from the last equation:
\begin{align}
	&p(\por | \data) = \varphi_{D^2} \left(\por; \boldsymbol{\mu_{\por|\dataRV}}, \boldsymbol{\Sigma_{\por|\dataRV}} \right), \\ 
	&\boldsymbol{\mu_{\por|\dataRV}} = \boldsymbol{\Sigma_{\por|\dataRV}} \left( 
	(\boldsymbol{J}_s \boldsymbol{J}_p)^T
	(\boldsymbol{\Sigma_{\dataRV}} + \boldsymbol{J}_s \ppesigma \boldsymbol{J}_s^T)^{-1}
	\data + \porsigma^{-1}~\pormu
	\right), \\
    &\boldsymbol{\Sigma_{\por|\dataRV}} = \left( \porsigma^{-1} + (\boldsymbol{J}_s \boldsymbol{J}_p)^T  (\boldsymbol{\Sigma_{\dataRV}} + \boldsymbol{J}_s \ppesigma \boldsymbol{J}_s^T)^{-1}  (\boldsymbol{J}_s \boldsymbol{J}_p) \right)^{-1}
\end{align}

For the case with linear physics, the importance density $\widetilde{p}(\latsmall | \por, \data) = \varphi_{L}(\latsmall; \muIS,\sigmaIS)$ introduced in Section \ref{S.IS} is an exact expression for $p(\lat | \por, \data)$ and the IS mean and covariance matrix reduce to:
\begin{align}
&\muIS~=~\sigmaIS \left( \boldsymbol{J_{s}}^T \boldsymbol{\Sigma_{\dataRV}}^{-1} \data + \ppesigma ^{-1} \mathcal{F}(\por) \right), \\ &\sigmaIS~=~(\ppesigma ^{-1} + \boldsymbol{J_{s}}^T \boldsymbol{\Sigma_{\dataRV}}^{-1} \boldsymbol{J_{s}})^{-1}. 
\end{align}

\end{document}